# Swift heavy ion track formation in SiC films under high-temperature irradiation

D.I. Zainutdinov[a*], A.E. Volkov[a]

[a]*P.N. Lebedev Physical Institute of the Russian Academy of Sciences, Leninskij pr., 53,119991 Moscow, Russia*

## Abstract

It is known that swift heavy ion (SHI) irradiation at temperatures below ~1000 K does not cause structural damage in the bulk SiC. However, the effect of the SiC film thickness on the formation and structure of SHI tracks over a wide range of irradiation temperatures remains unexplored. To address this gap, we used a model sensitive to irradiation temperature that describes all stages of ion track formation: from material excitation, considering the emission of excited electrons from the film surface (MC code TREKIS-3), to the reaction of the material's atomic system to the excitation (classical molecular dynamics). We observed the appearance of two different types of nanostructures on the surface of SiC films with thicknesses ranging from 10 nm to 100 nm after impacts of 710 MeV Bi ions: craters and hillocks. The transition from hillocks to craters occurs with the irradiation temperature. The transition temperature increases with the film thickness, tending to ≈1534±100 K for the surface of bulk SiC.

## 1. Introduction

Silicon carbide (SiC) is widely used in modern solid-state microelectronics and nuclear technologies due to its unique physical and chemical characteristics. Nowadays, there is a trend towards miniaturization of devices based on SiC [1], as well as their application in extreme conditions, which include both high temperatures and ionizing radiation [2–5]. In particular, it is of interest to study the response of SiC to irradiation with swift heavy ions (SHI) with masses larger than $10m_p$ ($m_p$ is the mass of a proton) and energies of 1-10 MeV/nucleon providing the ion stopping in matter in the electronic energy loss regime. Impacts of these ions can cause the failure of space SiC electronics due to the so-called single-event burnout phenomenon [6,7], or lead to the degradation of structural components in fission and fusion reactors [8].

*Corresponding author: d.zaynutdinov@lebedev.ru

Although some progress has been made in studying the effect of SHIs at high irradiation temperatures on bulk SiC crystals [9–12], the question of the influence of the size of these crystals on their response to irradiation with SHI beams remains open. Changes in crystal size can increase the role of the surface in the kinetics of defect formation as well as in the evolution of defects in the irradiated material [13–15] thereby significantly altering the response of the target to the irradiation.

In Ref. [16] we briefly reported on the damage to SiC films irradiated with SHIs at room temperature. However, the question of how this damage varies with the irradiation temperature remained open. Furthermore, many features regarding the shape and composition of the damage caused by SHI irradiation at room temperature were not investigated in Ref. [16]. This paper presents a detailed analysis of the formation kinetics of structurally altered SHI tracks in SiC films at different irradiation temperatures.

Several approaches exist to investigate structural changes in SHI-irradiated targets, including thermal spike model (or two-temperature model (TTM)) [17] and its modifications [18,19], as well as the Coulomb explosion model [20]. However, based on the conclusions drawn in works [21,22], we use a model more suitable for describing in detail the kinetics of fast, non-equilibrium processes in a SHI track.

We investigate structural changes in SiC films irradiated with SHIs at high temperatures using a well tested multiscale model that does not rely on macroscopic simplifications [11]. This model combines the Monte Carlo (MC) code TREKIS-3 [23,24], which describes the excitation of the electron and ion systems of the target over ∼100 fs, taking into account the irradiation temperature and the emission of excited electrons from the surface [25], with the LAMMPS molecular dynamics (MD) code [26] simulating the reaction of the atomic lattice to the excitation. Our simulations revealed formation of two types of surface nanostructures depending on the irradiation temperature: nanohillocks and nanocraters. The transition from hillocks to craters occurs with irradiation temperature. The transition temperature increases with the film thickness, tending to ≈1534 K for the surface of bulk SiC. Based on the simulation results for 710 MeV Bi ions, we propose a simple relationship linking the transition temperature with the film thickness.

In our work, we considered one of the most stable polytypes, 6H-SiC [27], which remains stable even at elevated temperatures [28]. On the other hand, our calculations allow us to draw conclusions that are common to all SiC polytypes.

# 2. Model

As in Ref. [16], the simulation of SHI track formation in SiC films is carried out in two steps. First, using the Monte Carlo code TREKIS-3, we simulate the passage of the SHI through the film, the generation of excited electrons and holes, and their subsequent evolution, taking into account interactions with the atomic and electronic subsystems of the target. At this step, we compute the spatial distribution of the energy density transferred to the lattice by the cooling time of electronic excitation over ~100 fs. This distribution forms the initial atomic velocities around the ion trajectory, enabling the subsequent description of structural changes in the lattice via classical molecular dynamics (LAMMPS code). In both MC and MD simulations, we take into account the irradiation temperature of the target.

## 2.1. Monte-Carlo simulation

The MC code TREKIS-3 describes the ionization of the target along the SHI trajectory, transport of valence holes and excited electrons including the emission of electrons from the film surface. TREKIS-3 models the scattering of electrons on the atomic lattice transferring kinetic energy to atoms as well as on target electrons resulting in the ionization of matter with the formation of secondary holes and electrons. This code also describes the Auger decay of holes accompanied by electron emission, radiative decay with photon emission, and further photon transport and absorption [23,24]. The code does not take into account interactions of an SHI with target atoms due to the negligible cross sections of this process in the range of ion energies considered in this paper.

The radial distribution of the energy density transferred to atoms in the vicinity of the ion trajectory forms the main result of the MC modeling. The energy is deposited into the lattice via two channels: (a) scattering of excited electrons and valence holes, as well as (b) nonthermal heating [21] resulting from an acceleration of atoms caused by a sharp change in the interatomic potential due to fast extreme excitation of the electron system of a target. Following the Ref. [29], nonthermal atomic heating can be taken into account by converting the potential energy of electron-hole pairs into the kinetic energy of atoms at the moment of cooling down of the electron system (approximately 100 fs after the ion impact) due to band gap collapse.

TREKIS-3 is based on the asymptotic trajectory method of event-by-event simulation, in which the scattering of a particle is described by cross sections that take into account the collective response of the target electrons and atoms to excitation in the linear response (first order Born) approximation [30].

The energy exchange between excited electrons/valence holes and the lattice is described using the non-ionizing scattering cross section. This cross section incorporates scattering on both phonons and isolated atoms and is expressed in terms of the atomic dynamic structural factor (DSF) $S_{at}(\omega, q, T_{irr})$:

$$\frac{d^2\sigma_{at}}{d(\hbar q)d(\hbar\omega)} = \frac{q m_{e,h}}{4\pi\hbar^4 E}\left|U_{p-at}(q)\right|^2 S_{at}(\omega, q, T_{irr}). \tag{1}$$

Here $\hbar\omega$ and $\hbar q$ are the energy and momentum transferred from the projectile to target atoms, and $\hbar$ is Planck's constant. $E$ is the projectile energy, $m_e$ is the electron mass, $m_h$ is the mass of a valence hole obtained in the one-band model [24] from the density of valence band states [31]. The interaction potential between a projectile (electron or valence hole) and a screened lattice ion is given by $U_{p-at,}(q) = 4\pi e^2/(q^2 + (l_{scr})^{-2})$, with the screening length $l_{scr}$=0.14 nm from Ref. [11]. The ion charges are included in atomic DSF and $e$ is the electron charge.

A change in the irradiation temperature leads to a change in the number of phonons in the crystal modes, the phonon lifetime, and their spectrum due to the expansion of the crystal lattice upon heating [32,33]. This leads to a change in the $S_{at}(\omega, q, T_{irr})$ and hence the scattering cross sections (1) on the atomic system. To account for the temperature dependence of the DSF, we recalculated the cross sections separately for each temperature using the algorithm described in Ref. [11]. In this approach, the classical atomic DSF ($S_{at}^{cl}(\omega, q, T_{irr})$), which is symmetric on transferred energy $\hbar\omega$, is calculated from the trajectories of target atoms obtained from the classical MD simulation at a fixed temperature. On the other hand, the quantum DSF $S_{at}(\omega, q, T_{irr})$ is asymmetric on the transferred energy: $S_{at}(\omega, q, T_{irr}) = \exp\,(\hbar\omega/T_{irr}) \cdot S_{at}(-\omega, q, T_{irr})$ and can be recovered from the classical DSF through the multiplication by the 'Harmonic' quantum correction factor [34–36]:

$$S_{at}(\omega, q, T_{irr}) = \frac{\hbar\omega/k_b T_{irr}}{1 - exp(-\hbar\omega/k_b T_{irr})} S_{at}^{cl}(\omega, q, T_{irr}) \tag{2}$$

The obtained quantum atomic DSF satisfies the quantum sum rule [11] and contains temperature in two parts: in the quantum correction factor and in the classical DSF, where the temperature is present in the trajectories of atoms in the form of their velocity distribution.

The particle scattering cross section (SHI, electrons, valence holes) on the electron subsystem is conveniently expressed through the imaginary component of the electron part of the inverse complex dielectric function (CDF) of a material [37] – the loss function (LF), which weakly changes with the irradiation temperature:

$$\frac{d^2\sigma_e}{d(\hbar q)d(\hbar\omega)} = \frac{[Z_{eff}(v)e]^2}{\hbar^2\pi E n_e}\frac{m_{e,h}}{\hbar q}\left(1 - e^{-\frac{\hbar\omega}{k_b T_{irr}}}\right)^{-1} Im\left[-\frac{1}{\varepsilon_e(\omega,q)}\right], \tag{3}$$

Here $n_e$ is the concentration of electrons in the medium, and $Im[-\varepsilon_e^{-1}(\omega,q)]$ is the electron part of the LF, which is presented for SiC in Ref. [11]. The scattering cross section on the electron subsystem of a material accounts for ionization of the atomic shells and valence band. The effective charge of an SHI, $Z_{eff}(v)$, is estimated using the Barkas formula [21,38]:

$$Z_{eff}(v_{ion}) = Z_{ion}\left[1 - exp\left(-\frac{v_{ion}}{v_0}Z_{ion}^{-\frac{2}{3}}\right)\right], \tag{4}$$

where $Z_{ion}$ is the nuclear charge of the incident ion, $v_0 = \mathrm{c}/125$, where $c$ is the speed of light in vacuum. The effective charges of electrons and valence holes are equal to unit.

The transition probability of electrons from the film surface to vacuum in TREKIS-3 [25] is determined by Eq. (5), which takes into account their reflection and the effect of quantum tunneling through the surface barrier:

$$\tau(E) = \frac{1}{1 + \exp\big(\gamma(E_1 - E)\big)} \tag{5}$$

Here $E$ is the electron energy, $E_1$ is a parameter that depends on the height of the emission barrier, and $\gamma$ is a parameter that depends on the characteristic spatial scale of the barrier [39]. The energy of the emitted electron is counted from the vacuum energy level (kinetic energy of the electron minus the work function). The parameters $E_1$ and $\gamma$ depend on the substance. The procedure for determining them for SiC as well as the corresponding Eckart-type barrier are described in the next section.

### 2.2. Transmission coefficient and emission barrier in SiC

For the transition from a sample to a vacuum, an electron must overcome a potential barrier. We applied Eckart's form of this barrier $V(z)$, which allows for an analytical calculation of the transmission coefficient [40]:

$$V(z) = -\left[\frac{W\xi}{1-\xi} + \frac{B\xi}{(1-\xi)^2}\right] \tag{6}$$

here $\xi = -\exp{(2\pi z/L)}$, $z$ is a Cartesian coordinate. Negative values of $z$ determine the potential inside the target, where $V(z \ll -L) \approx 0$, and positive values on the surface $V(z \gg L) \approx W$ with $W$ being the work function. $L$ is the characteristic length of the barrier (width of the transition region is $2L$) and $W_1$ is its height determined by the work function and the parameter $B$:

$$W_1 = \frac{(W+B)^2}{4B} \tag{7}$$

Although the Eckart-type barrier was originally developed to describe thermionic emission process, a number of studies have demonstrated its applicability for describing emission in SHI tracks. For example, Ref. [25] showed that using an Eckart-type barrier to describe electron emission during SHI irradiation of semiconductors leads to good agreement with the experimental spectrum of emitted electrons. In addition, references [13,41] showed that the track structures calculated with this barrier are consistent with the experimental data. Ref. [42] demonstrated the agreement between the calculations performed with the Eckart-type barrier and the results of more complex models, which, in particular, allow for the consideration of the attraction between escaped electrons and holes remaining in the track. Taking this into account, we assume that the Eckart-type barrier is suitable for solving our problem.

The simple formula (5) can be used for the transmission coefficient $\tau(E)$ for a narrow Eckart-type barrier, when $W_1 > W$ [39]. Following this work, we determined the parameters $E_1$ and $\gamma$ by fitting the experimental spectrum of emitted electrons [43] with the function:

$$\tilde{J}(E) = \frac{(E+W)^{\frac{1}{2}}\left[C_1 \exp\left(-\frac{E}{\epsilon}\right) + C_2 \exp\left(-\frac{E}{2\epsilon}\right)\right]}{1+\exp\bigl(\gamma(E_1 - E - W)\bigr)} \tag{8}$$

where $\tilde{J}(E)$ is the number of electrons emitted into the vacuum from a unit of surface, per unit time, per unit energy; $\epsilon$ is the mean energy exchange between the secondary electrons and the medium at the free path length; $C_1$ and $C_2$ are related to the intensity of the $\epsilon$ and $2\epsilon$ contributions to the distribution, respectively.

Table 1 shows the values of the parameters in Eq. (8) that provide good agreement with the experimental data on thermionic emission (see Fig. 1a) [43] from the (001) surface of 6H-SiC crystal. The value of the work function $W = 4.5$ eV was taken from Ref. [44].

Table 1. Values of parameters of function (8) realizing the best agreement with the experimental data [43] on thermionic emission of electrons from 6H-SiC .

| $W, eV$ | $C_1$ | $C_2$ | $\epsilon, eV$ | $\gamma, eV^{-1}$ | $E_1, eV$ |
|---|---|---|---|---|---|
| 4.5 | 61.2 | 5.3 | 1.58 | 8.48 | 4.68 |

We obtained the parameters $L = 1.2$ nm and $B = 6.97$ eV of the Eckart-type barrier corresponding to the experimental values of $E_1$, $\gamma$, $W$ from Table 1 using the system of equations (7-8) from Ref. [39].

Fig. 1b shows the transmission coefficient through the barrier which corresponds to the parameters $E_1$ and $\gamma$ from Table 1 and takes into account the reflection of electrons with energies $E < W$.

Fig. 1c shows the dependence of the Eckart-type barrier in SiC on the distance $z$ to the surface, illustrating that the assumption $W_1 > W$ is valid for this barrier. The difference between the barrier height $W_1 = 4.72$ eV and the work function $W = 4.5$ eV occurs due to the existence of an image potential [39] in SiC.

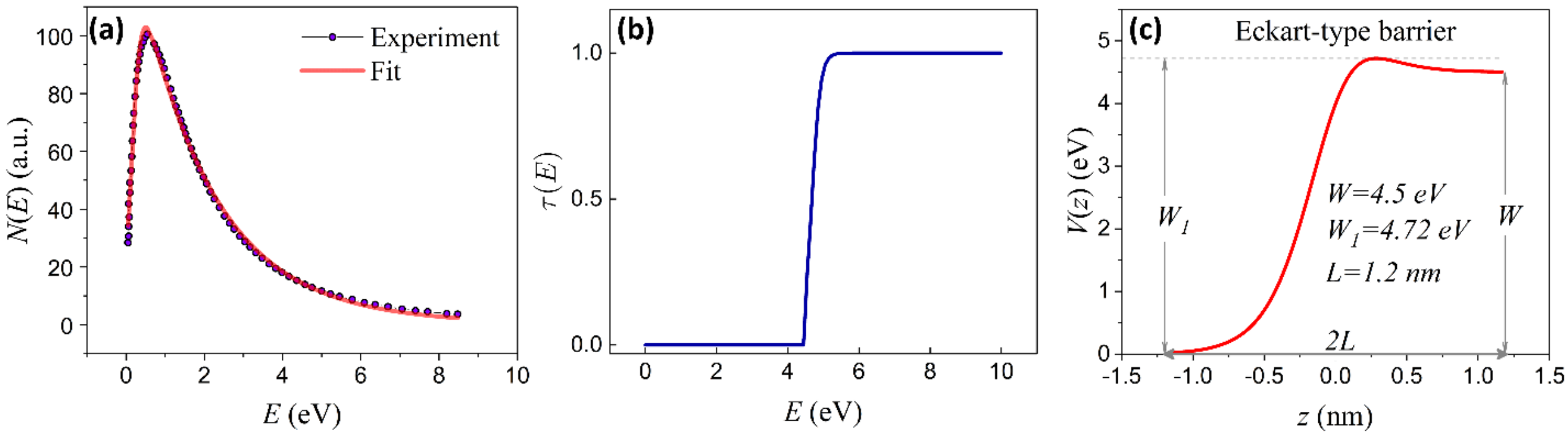

*Fig. 1 (a) Experimental distribution of thermal-emitted electrons from the SiC surface* [43] *and its fitting with Eq. (8); (b) Transmission coefficient (5) through the surface potential barrier in SiC; (c) Shape of the Eckart-type barrier (6) in SiC*

The work function $W$ is the main parameter determining electron emission from the film surface. Taking into account previous conclusions, we determine the corresponding barrier height as $W_1 = W + 0.22$ eV. According to Ref. [45], the work function in SiC remains unchanged within the experimental accuracy as the temperature increases up to 1500 K and is equal to 4.6±0.1 eV , which coincides well with the value of $W = 4.5$ eV used in our work. Appendix A presents an additional analysis of the work function at temperatures ranging from 1500 K to 2200 K. It shows that the temperature dependence of the work function in this range can also be neglected. Therefore, in all the simulations, we consider the barrier to be constant.

### 2.3. Molecular-dynamics model

We prepared rectangular parallelepiped-shaped films with thicknesses ranging from 10 nm to 100 nm containing 6H-SiC crystals with crystal axes $[10\bar{1}0]$, $[\bar{1}2\bar{1}0]$, and $[0001]$. With this orientation, the simulation cells had a C-terminated (0001) surface on top and an Si-terminated (0001) surface on the bottom [46]. In all simulation cells, except the one containing the 100 nm thick film, free boundary conditions were applied at a distance of 30 nm from the film surfaces along the SHI trajectory direction [0001]. For the 100 nm thick film, the free boundaries were placed 50 nm from the surfaces. Periodic boundary conditions were used in the $[10\bar{1}0]$ and $[\bar{1}2\bar{1}0]$ directions. The dimensions of the films in the (0001) plane are $28 \times 29$ nm$^2$.

We used the Vasishtha's potential [47] to describe interatomic interactions. We chose this potential because, unlike other MD potentials created for SiC, it correctly reproduces the recrystallization of the lattice in the SiC volume after irradiation with 710 MeV Bi ions at room temperature [11] leading to agreement with the experimental data [48]. It correctly reproduces the main processes of structural changes in the excited and hot track region: disorder, material transport, and recrystallization.

Prior to simulations, we minimized the energy in the films using the selected potential and performed a two-step equilibration procedure at a fixed temperature, as described in Ref. [49].The kinetic energy for each atom in the track was determined randomly according to a Gaussian distribution of the atomic velocities. To build up this distribution, we used the average energy density released in the atomic lattice at the atomic position by the time of electron cooling, taken from the TREKIS simulation at ~100 fs after the ion impact [49].

For the MD simulations, the initial velocities of atoms in the ion track were generated using the algorithm described in Ref. [50,51]. The algorithm employs the radial distribution of the energy density (energy per atom) deposited into the atomic lattice around the ion trajectory over the characteristic relaxation time of the electronic subsystem in the track (100 fs). Depending on the distance to the ion trajectory, the average kinetic energy of atoms was assumed as the sum of the deposited energy at this distance and the thermal energy corresponding to the specified irradiation temperature. For each individual atom at a given radial distance, the kinetic energy was assigned randomly according to a Gaussian distribution centered on this average value. The direction of the velocity vector was also chosen randomly following a uniform distribution.

Since the algorithm generates the initial velocities of atoms stochastically, additional verification of energy conservation was required. Therefore, for two simulation cells (the thinnest, 10 nm, and the thickest, 100 nm), we compared the total kinetic energies, minus the irradiation temperature, with the expected values. The total kinetic energy in the MD cell was calculated using equation: $E_{at}^{MD} = \sum_{i=1}^{N_{at}} m_i v_i^2/2 - \frac{3}{2} k_B T_{irr} N_{at}$, where $T_{irr}$ is the irradiation temperature, $m_i$ and $v_i$ are the mass and velocity of $i$-th atom, respectively, and the summation is performed over all atoms in the cell. The expected energy was also computed based on the profile of energy transferred to the lattice during 100 fs from TREKIS-3, without using a random number generator. This involved summing the energies assigned to each atom according to its distance from the ion trajectory. The results of the comparison are presented in Table 2.

Table 2. Comparison of the total energies deposited into the lattice after irradiation with a 710 MeV Bi ion in MD cells for 10 nm and 100 nm thick films at three irradiation temperatures. The values are given both without velocity randomization (lattice energy from TREKIS-3) and with it (lattice energy in MD)

| Film thickness (nm) | Number of atoms | $T_{irr}$ (K) | Lattice energy form TREKIS (keV) | Lattice energy in MD (keV) |
|---|---|---|---|---|
| 10 | 797,040 | 300 | 280.48 | 280.46 |
| 10 | 797,040 | 1000 | 247.06 | 247.02 |
| 10 | 797,040 | 2200 | 250.87 | 250.92 |
| 100 | 7,678,800 | 300 | 2653.63 | 2653.99 |
| 100 | 7,678,800 | 1200 | 2845.35 | 2845.67 |
| 100 | 7,678,800 | 1600 | 2886.12 | 2886.58 |

Several conclusions can be drawn from Table 2. First of all, not all the energy lost by the ion (≈35 keV/nm multiplied by the film thickness) is converted into the energy of atoms near the ion trajectory, within a distance corresponding to the size of the MD cell (several tens of nanometers). Part of the energy in MC simulations is deposited into the lattice at large distances ∼100 nm or lost in the process of electron emission. Second, energy conservation is satisfied in the transition from MC to MD simulations. The largest difference between the output MC data and the input MD data is observed in the 10 nm cell at a temperature of 2200 K and is approximately $6\times10^{-8}$ eV/atom, which is much less than the thermal fluctuations at a given irradiation temperature.

We modeled the evolution of the atomic structure after the SHI passage in the LAMMPS code using Vashishta's interatomic potential [47] and the Verlet algorithm in a simulation cell that includes a film with open surfaces. We used a variable time step corresponding to a maximum atomic displacement of 0.1 Å. The modeling finished when the track core cooled to the irradiation temperature (~200-300 ps). During the simulation, the Berendsen thermostat [52] with a damping time of 20 fs was applied in a 5 Å thick layer at the cell boundaries around the ion trajectory.

# 3. Results and Discussion

## 3.1. Excitation of SiC film with electron emission

Ref. [13] demonstrated that the effect of electron emission on ion track formation is most significant in thin films up to 15 nm thick. Therefore, to investigate the influence of irradiation temperature on the excitation of the electron system of a target taking into account electron emission, we performed a series of MC simulations of irradiations of a 10 nm thick SiC film with Bi 710 MeV ion ($dE_e/dx \approx 35$ keV/nm) at different temperatures.

For the transition of excited electrons from the film surface to vacuum, two conditions must be satisfied:

1. The electron velocity component normal to the film surface directed towards the vacuum must be nonzero [39];
2. The electron energy must be greater than the work function (see Fig. 1b).

It means that the number of emitted electrons and, consequently, the profile of the energy deposited in the atomic lattice at different irradiation temperatures are determined by, at least, two processes: the isotropization of excited electron momentum and their deceleration in the material.

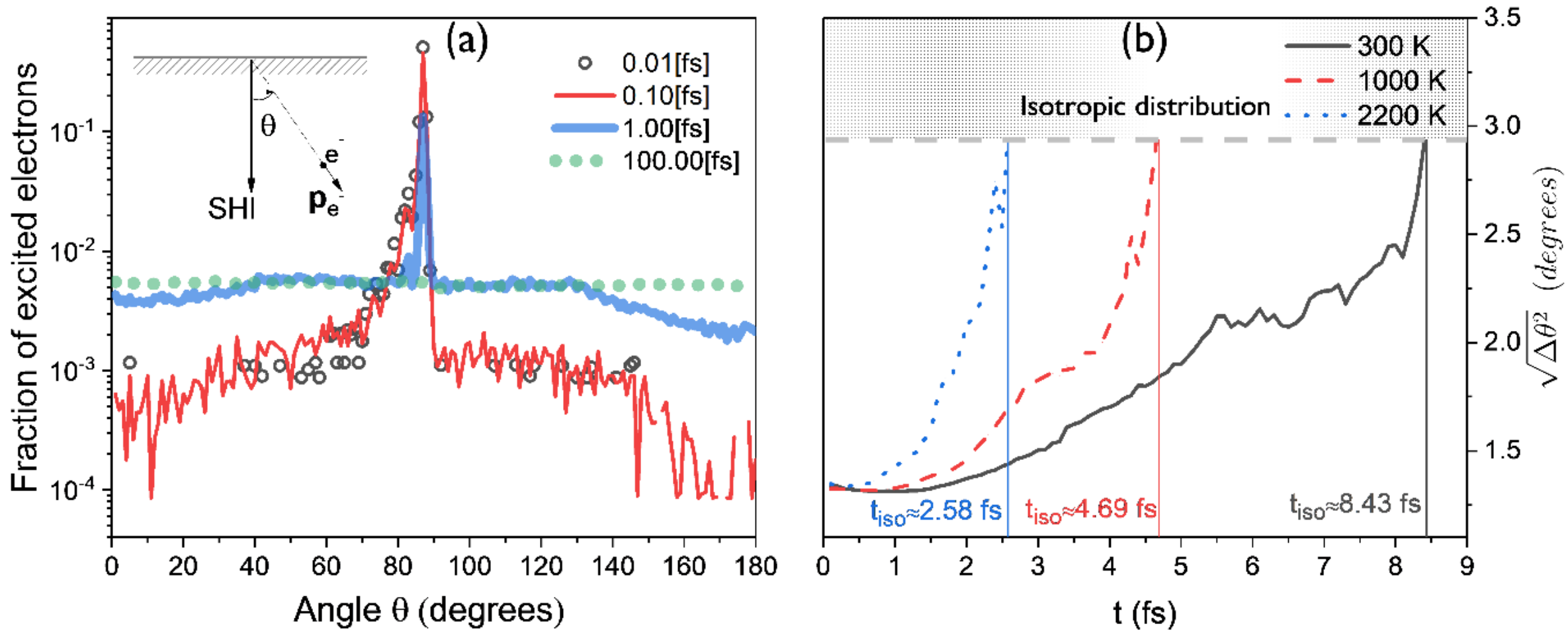

*Fig. 2. (a) Momentum distribution of excited electrons along the zenith angle θ following irradiation of a 10 nm thick SiC film by a 710 MeV Bi ion at 300 K. (b) Temporal evolution of the full width at half maximum at θ=90° of the excited electron angular distribution in a 10 nm thick SiC film irradiated by 710 MeV Bi ion at different temperatures.*

The passage of an SHI through the film perpendicular to its surface leads to the formation of a $\delta$ -electron front after ~0.01 fs, propagating predominantly perpendicular to the ion trajectory and parallel to the surface. Consequently, the angular distribution of excited electrons relative to the zenith angle $\theta$ (see Fig. 2a) at the moment of 0.01 fs exhibits a pronounced peak centered at about $\theta \approx 90°$, where $\theta = 0°$ corresponds to the electron motion along the ion direction, and $\theta = 180°$ - in the opposite direction. The small variation in the momentum direction of these electrons leads to the possibility that a small fraction of them may escape the film within 0.1 fs after irradiation, as can be seen in Fig. 3a. The fraction of escaped electrons at these times slowly increases with the irradiation temperature due to the increase in the number of excited electrons in the track, as shown in Ref. [11]. However, this trend is very weak and has little effect on the final profile of electrons escaped over 100 fs at different temperatures.

Starting at 0.1 fs, the scattering of electrons on the atomic lattice leads to isotropization of their zenith angle distribution. Fig. 2a shows how the sharp peak at 90° at 0.1 fs in the distribution of excited electrons formed at room irradiation temperature broadens and decreases in height, passing to a uniform distribution by 100 fs. As a result, some electrons leave the film during the isotropization process.

Moreover, increasing the irradiation temperature leads to an increase in the isotropization rate, as shown in Fig. 2b. For example, at $T_{irr} = 300$ K electrons are isotropized during the time $t$ ≈8.4 fs, at $T_{irr} = 1000$ K during $t$ ≈4.7 fs, and at $T_{irr} = 2200$ K during $t$ ≈2.6 fs. Here we used the width of the peak at half-height in the electron zenith angle $\theta$ distribution, denoted by $\sqrt{\Delta\theta^2}$, as a measure of the distribution's isotropy in Fig. 2b.

The polynomial fit of the data in Fig. 3a shows an increase in the number of emitted electrons from the SiC film 100 fs after irradiation at temperatures ranging from 300 K to 500 K. This increase can be associated with a rise in the isotropization rate of excited electrons, whereby electrons with above-barrier energy reach the surface before losing their energy due to lattice scattering.

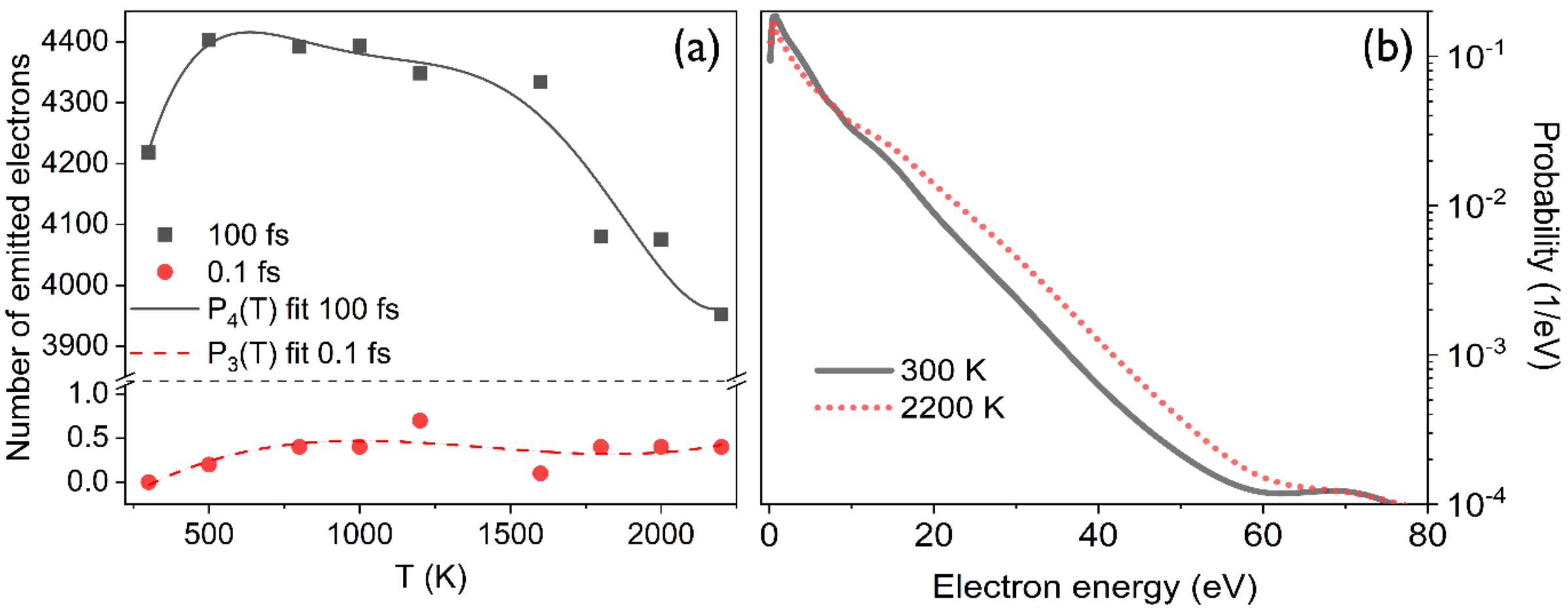

*Fig. 3 (a) Total number of emitted electrons from a 10 nm thick SiC film at 0.1 fs and 100 fs after Bi 710 MeV ion irradiation at different temperatures averaged from 10 MC simulations and their approximations by third- and fourth-degree polynomials, respectively; (b) Normalized kinetic energy spectra of secondary electrons emitted from a 10 nm thick SiC film after Bi 710 MeV ion impact at 300 K and 2200 K*

However, starting from $T_{irr} = 500$ K, as shown in Fig. 3a, increasing the irradiation temperature leads to a decrease in the number of emitted electrons. This decrease can be associated with an increase in the energy loss of excited electrons per unit path caused by their scattering on phonons (see Ref. [11]). A similar trend was observed experimentally [53] in the range of 463-659 K when the bulk SiC target was irradiated with $Xe^{17+}$ ions with energies of 3.2 MeV and 4 MeV. Apparently, this trend at high irradiation temperatures is not associated with a change in the work function, which practically does not change with temperature [44], but is determined by the scattering of excited electrons on the atomic system [54].

To investigate the effect of electron emission on the profile of energy deposited in the atomic lattice in an SHI track, we performed a series of MC simulations of a 10 nm SiC thin film irradiated with a 710 MeV Bi ion at various temperatures: (i) including electron emission and (ii) with periodic boundary conditions (bulk simulations).

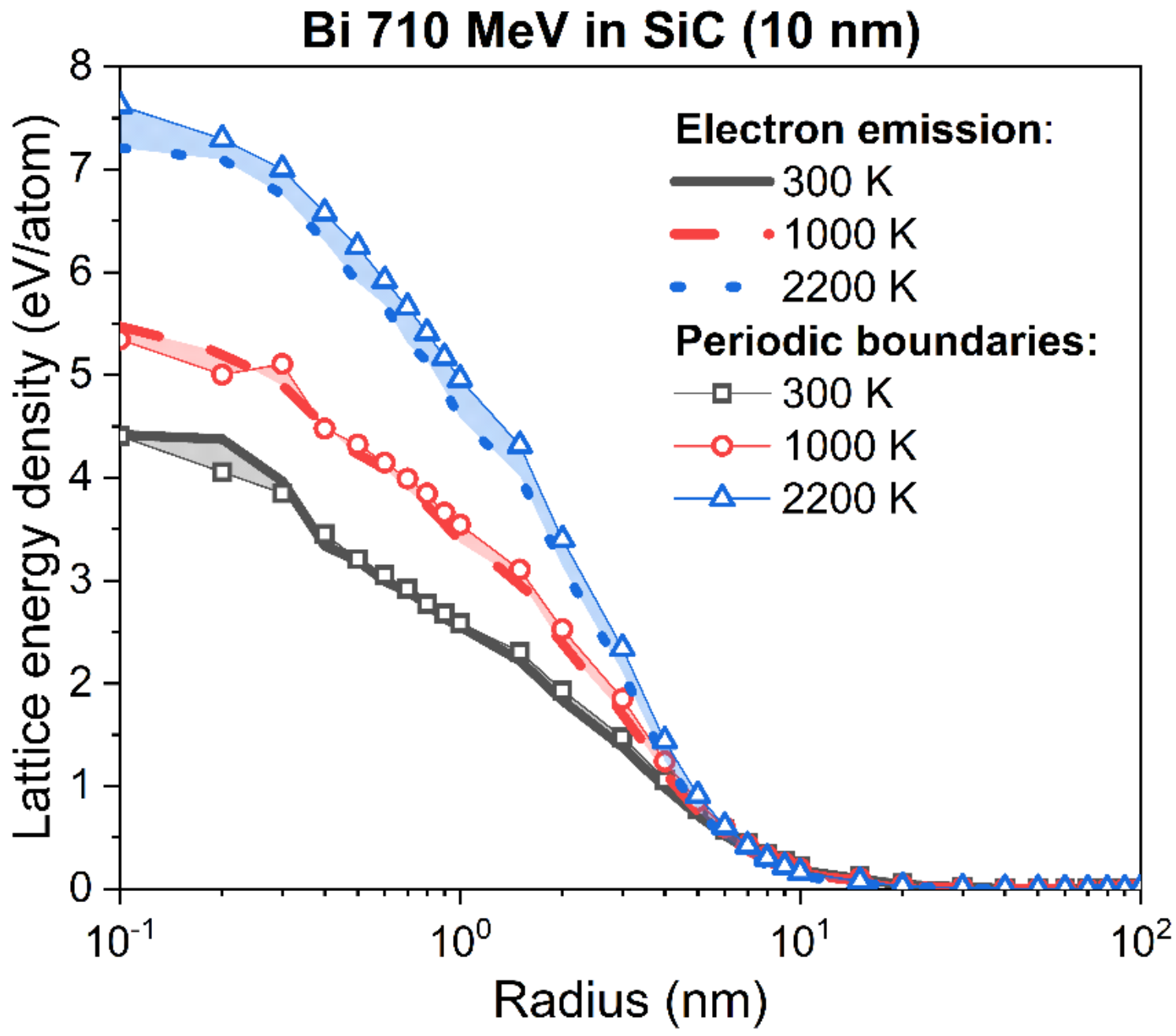

*Fig. 4 Profiles of the energy transferred to the atomic lattice of 10 nm thick SiC film 100 fs after irradiation with a 710 MeV Bi ions at different temperatures. The color indicates the difference between the profiles obtained at the same temperature with and without electron emission .*

Fig. 4 shows the radial distributions of the energy density transferred to the lattice in the tracks of 710 MeV Bi ions at different irradiation temperatures for these two cases. Evidently, the influence of emission on the deposited energy profile is noticeable only at high irradiation temperatures, at $T_{irr} \approx 2200$ K in our case. In the simulations with emission, less energy is transferred to the lattice than in the simulations with the periodic boundary conditions. However, Fig. 3a shows that at these high temperatures (for example $T_{irr} \approx 2200$ K) the number of emitted electrons is much smaller than their number at lower irradiation temperatures. As a result, increasing the irradiation temperature suppresses the process of electron emission, but at the same time increases the role of electron emission in the formation of the final profile of the energy transferred to the atomic system.

Qualitatively, it can be associated with an increase in the fraction of phonon absorption in the scattering of electrons on the atomic lattice, which increases with irradiation temperature due to a decrease in the asymmetry of the atomic DSF in terms of transferred energy (see Eq. (2)). As a result, Fig. 5 shows that the rate of energy transfer to the lattice decreases with irradiation temperature. However, its temporal evolution remains unchanged: increasing the time up to about 20 fs is associated with the formation of “slow” electrons in cascades, which are predominantly scattered on the lattice, and a further decrease associated with the cooling of the electronic subsystem.

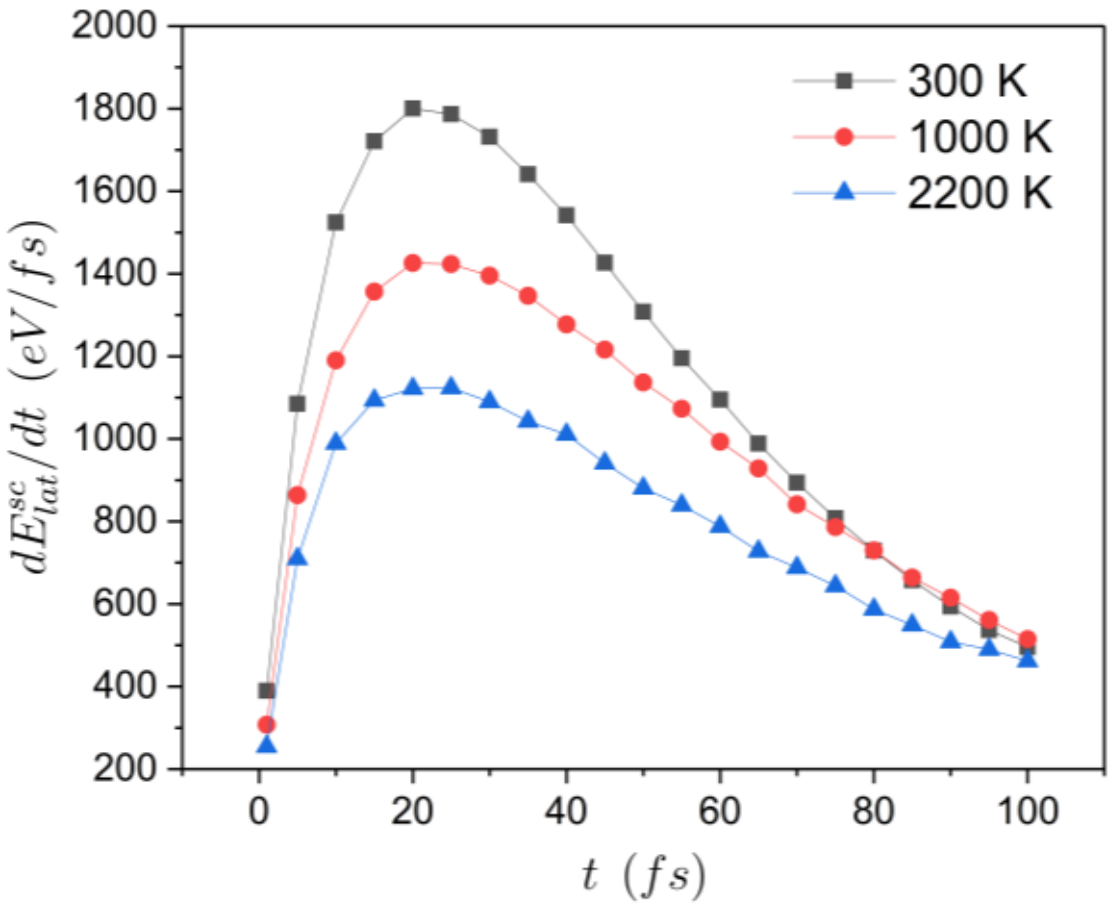

*Fig. 5 The energy transfer rate to the entire atomic lattice of a 10 nm thick SiC film through the scattering of excited electrons in the track of 710 MeV Bi ions at different irradiation temperatures, taking into account electron emission.*

Due to the decrease of the energy transfer rate, at high irradiation temperatures hot electrons escape the film before transferring all their energy to atoms. This leads to less heating of the film compared to the simulations with the periodic boundary conditions. Fig. 3b illustrates the normalized kinetic-energy distribution function of the emitted electrons, where the fraction of high-energy electrons increases with the irradiation temperature. However, the difference between the profiles of the energy deposited to atoms, obtained taking into account the emission and without it, is small (~8 % , see Fig. 4) and, as shown in Ref. [13], becomes smaller with increasing film thickness. Furthermore, based on the additional analysis of these profiles presented in Appendix A, two conclusions can be drawn. First, the difference between them does not affect the structural changes in the track that are of interest to us in this work. Second, this difference may be even smaller if we take into account the temperature dependence of the work function. Based on these conclusions, for films thicker than 30 nm, we neglected electron emission and performed the MC simulation with periodic boundary conditions, which allowed for a reduction in computational costs.

### 3.2. Structural changes in a 10 nm thick SiC film

Fig. 6a shows the temporal evolution of the atomic structure in a 10 nm thick SiC film after 710 MeV Bi ion impact at room temperature. It can be seen that during 5 ps, a molten cylindrical column with a diameter of $d$ ≈7 nm is formed in the track center. The SiC melt has the lower density than the crystalline phase [11]. Therefore, thermal expansion of the molten region leads to "the squeezing" of some substance onto the film surface.

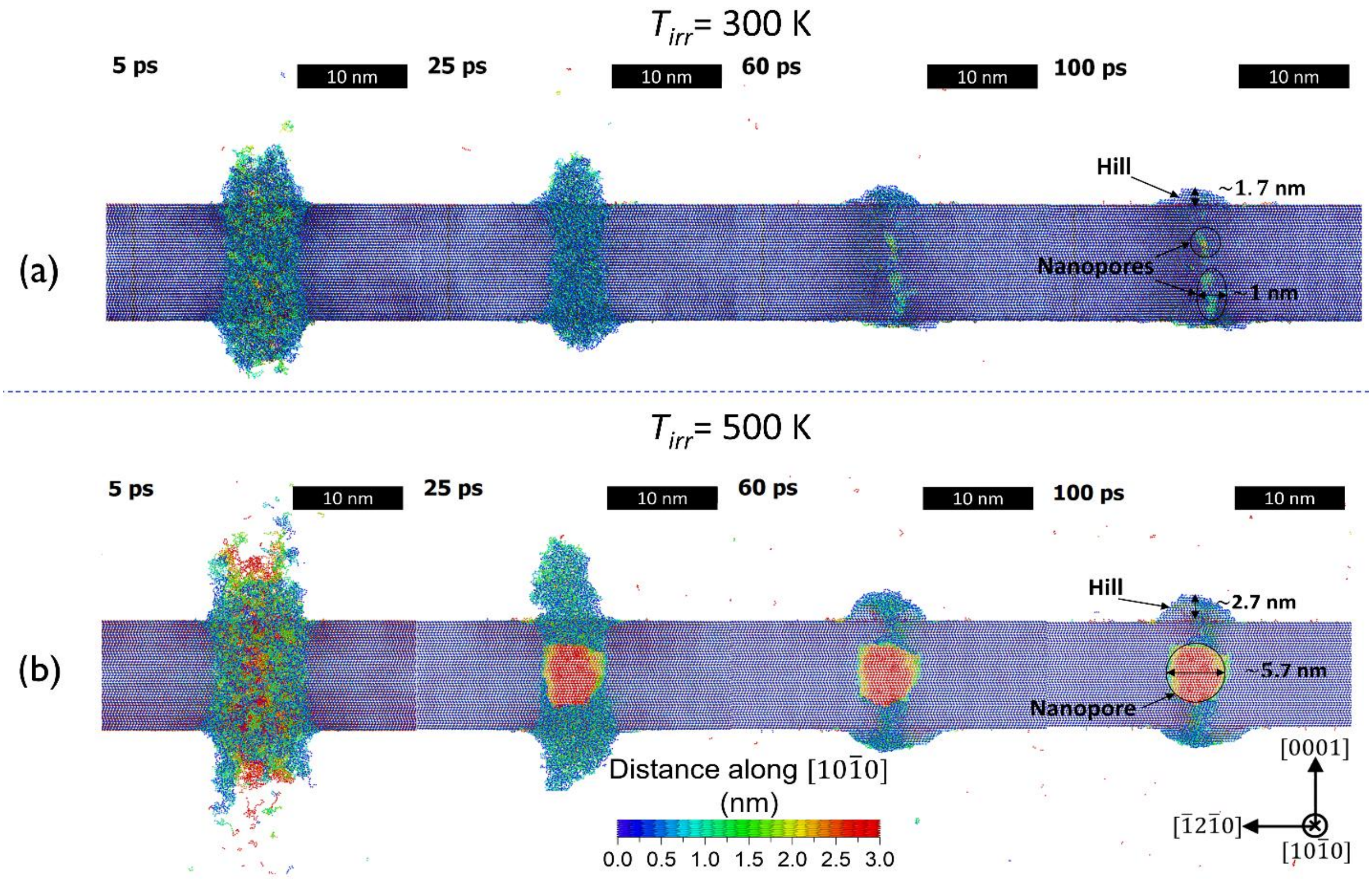

*Fig. 6 Temporal evolution of the atomic structure of 10 nm thick SiC film after 710 MeV Bi ion imact (a) at 300 K and (b) at 500 K. The distance from the lattice atoms to the ion trajectory along the direction* $[10\bar{1}0]$ *is highlighted in color.*

Over the next 95 ps, the molten region recrystallizes. The similar recrystallization was observed in theoretical [50] and experimental [55,56] studies. The recrystallization rate is faster in the layers farthest from the surfaces. For example, Fig. 6a illustrates that at time 25 ps, the molten region has an hourglass shape. Starting from 35 ps, dislocations are formed at the edges of the disordered region near the film surface (see Fig. 7a). By 60 ps after irradiation, the track inside the plate is almost completely restored, but the surface hillocks still have a disordered structure (see Fig. 6a). Then, over the next 40 ps, the hillocks also recrystallize, with the recrystallization front moving away from the film surface. As a result, hillocks remain on the film surfaces at 100 ps after irradiation. The fraction of the crystalline state in them is ≈12%. The highest hillock has $h$ ≈1.7 nm height and $d \approx 6$ nm diameter. Nanocavities with a diameter of $d_c \approx 1$ nm surrounded by dislocations (see Fig. 7a) remain in the film volume along the ion trajectory. The surface of the nanocavities has a disordered structure consisting of point defects and defect clusters.

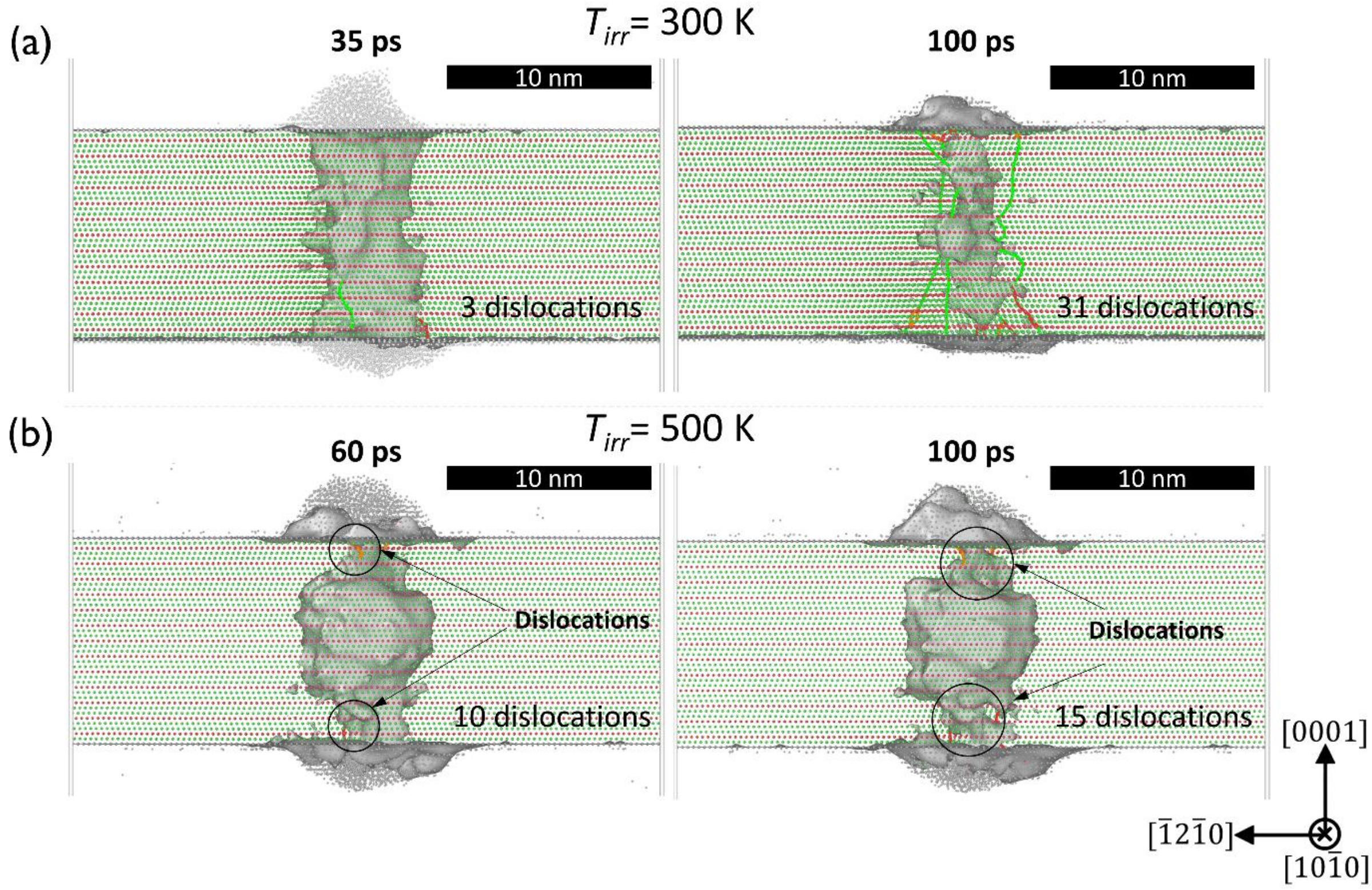

*Fig. 7 Carbon sublattice of a 10 nm thick SiC film with highlighted dislocation lines and disordered region at different times after impacts of 710 MeV Bi ions (a) at 300 K and (b) at 500 K.*

The irradiation of the same film with 710 MeV Bi ion at 500 K initiates similar stages of track formation (see Fig. 6b). However, dislocations form later: starting from 60 ps (see Fig. 7b). The main effect of the irradiation at 500 K manifests in the sizes of the residual defects: the height of the largest hillock is $h \approx 2.7$ nm, its diameter is $d \approx 7.6$ nm and the diameter of the inside cavity is $d_c \approx 5.7$ nm. The higher irradiation temperature also changes the defect composition: the crystallinity of the hillock increases to 31.4%, the surface of the cavity also becomes crystalline. The dislocations at the edges of the cavity remain only near the film surface (see Fig. 7b), and their number also decreases with increasing irradiation temperature. The observed intense recrystallization at high irradiation temperatures is consistent with the conclusions of a number of experimental studies [9,10,57].

Note that the dislocations in Fig. 7, which we observed in the track at different irradiation temperatures, do not move away. As a result, they do not participate in the process of nanocavity formation by the mass transfer from the track core, as was observed in the bulk simulations [11]. Therefore, the mechanism of formation of nanocavities inside the film near the surface is associated with the displacement of a part of the substance to the film surface or ejection of substance droplets, as will be shown further at higher irradiation temperatures.

Fig. 8 shows a 10 nm thick SiC film after atomic structure relaxation (200 ps after an ion impact) in a slice along the trajectory of 710 MeV Bi ion at different irradiation temperatures.

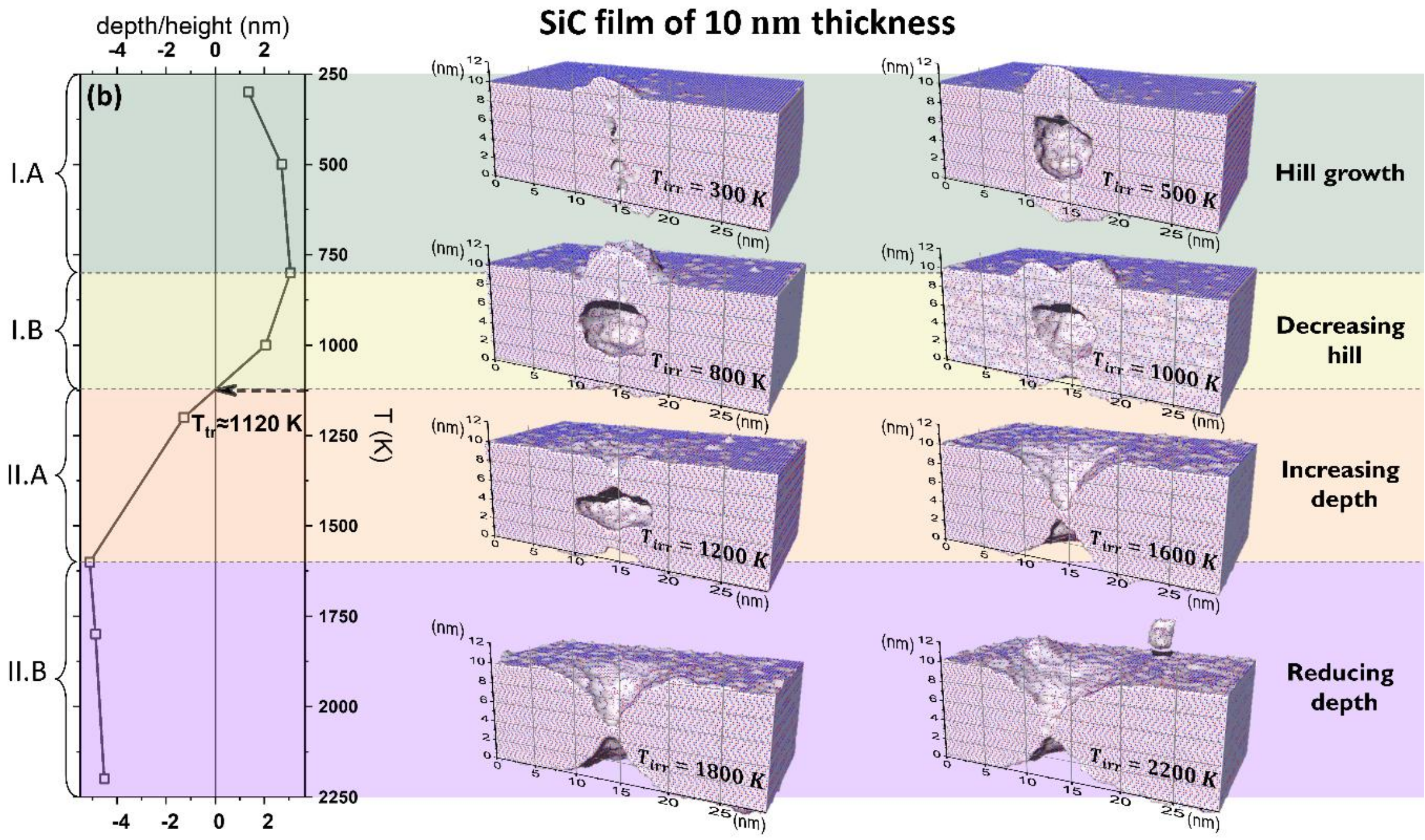

*Fig. 8 10 nm thick SiC film 200 ps after impacts of 710 MeV Bi ion at different temperatures, as well as the relation between the hillock height and crater depth for this film as a function of irradiation temperature.*

The presented results show that the temperature scale can be divided into four regions, which are illustrated in Fig. 8:

**I.A** Hillocks form on the film surfaces at irradiation temperatures up to 800 K. Their height increases with the irradiation temperature. At the same time, nanocavities remain in the film volume along the ion trajectory. Their volumes also increase with the irradiation temperature.

**I.B** An increase in the irradiation temperature in the range from 800 K to 1120 K leads to a decrease in the height of the hillocks, while the cavity sizes do not change;

**II.A** At the irradiation temperature ranging from 1120 K to 1600 K, the track geometry changes: a hillock transforms into a crater. The temperature increase lead to a decrease in the size of the cavity inside the film and an increase in the crater depth;

**II.B** At the irradiation temperatures above 1600 K, craters connected by a thin bridge remain on the film surfaces (see Fig. 8). In this case, the track does not contain cavities. An increase in the irradiation temperature leads to a thickening of the bridge and, consequently, a reduction in the crater depth. At the same time, the crater walls become more and more gentle. This effect is likely due to enhanced atomic mobility at higher temperatures, which promotes the smoothing of density inhomogeneities.

Unfortunately, experimental data for the irradiation temperature regimes above a thousand Kelvins are currently unavailable. Moreover, the boundaries of the detected temperature ranges may depend on both the ion energy (see Appendix B for more details) and the thickness of the target under consideration, thus requiring individual estimation for each specific experiment.

On the other hand, a number of experimental studies [58,59] have observed temperature regime I.A, where the growth of nanohillocks occurred with increasing irradiation temperature. Ref. [58] investigated the damage formation in the near-surface region of SiC after irradiation at 700 and 1000 K with He ions with energies up to 100 keV. These ions penetrate SiC to a depth of up to 500 nm and their electronic energy loss predominates over the nuclear energy loss in the near-surface region. The experiment shows the formation of helium-filled bubbles near the SiC surface after irradiation. The size of the bubbles increases with the irradiation temperature. This effect is similar to what we observed in our simulations (see Fig. 8). However, nanocavities detected in [58] are subsequently filled with He ions, which may affect their characteristics and require separate, more detailed consideration. Similar conclusions were drawn in Ref. [59], although the experimental conditions differed: SiC was irradiated with 30 keV He ions at 1073.15 K.

When determining the corresponding temperature regimes in Fig. 8, we used the height/depth relation of hillocks/craters averaged over both sides of the film. However, as can be seen, for example, in Fig. 6, the sizes of the nanohillocks on different sides of the film are different: the upper nanohillock is approximately 40% larger than the lower one. With increasing irradiation temperature, this difference rapidly decreases and becomes practically negligible in the case of craters. We think this effect is associated with the structure of our supercells, which feature the top layer composed of light carbon atoms (C-terminated surface) and the bottom layer of heavier silicon atoms (Si-terminated surface) [46].

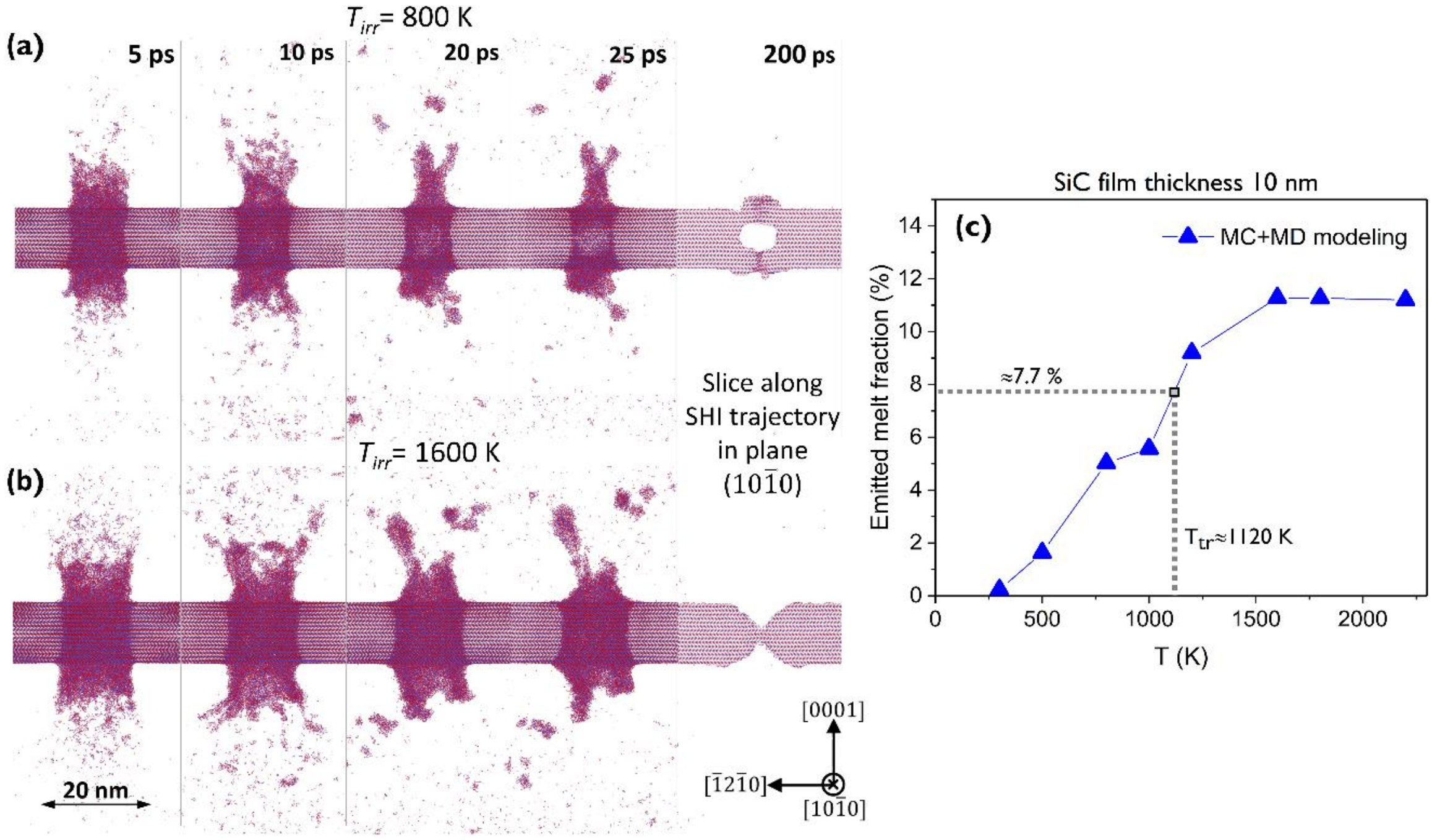

*Fig. 9 Evolution of a 10 nm-thick SiC film after impact of a 710 MeV Bi ion (a) at 800 K and (b) at 1600 K; (c) fraction of ejected melt from a 10 nm-thick film consisting of 797040 atoms as a function of irradiation temperature.*

Fig. 8 shows that nanohillocks transform into craters when the temperature exceeds $T_{tr} \approx 1120$ K. This transition is associated with the increase in the density of energy deposited into the lattice within the track at higher irradiation temperatures, as noted in Section 3.1. At temperatures $T_{irr} < T_{tr}$, the process of melt displacement to the film surface in the track region dominates (see Fig. 9a), leading to the formation of hillocks. At higher temperatures $T_{irr} > T_{tr}$ it is replaced by the process of ejection of matter droplets (see Fig. 9b), which leads to crater formation. The crater formation needs ejecting more than ~7.7% of the melt from the track (see Fig. 9c). This fraction may vary depending on the film thickness. The experimental observation of a similar transition under irradiation of LiF crystal with Xe ion is presented in Ref. [60], where an increase in local heating in the track was achieved by increasing the charge of the Xe ion from +15 to +36 (in units of electron charge).

The transition temperature from hillocks to craters can be significantly affected by the ion incident angle. As noted in Ref. [41], under oblique irradiation, most of the excited electrons predominantly move perpendicular to the ion trajectory towards the film surface. The reflection of slow electrons and valence holes from the surface leads to enhanced heating of the near-surface layer [41]. This makes the formation of craters possible even at room temperature. Ref. [61] illustrates the formation of nanochannels with a depth of ≈0.2 nm in SiC irradiated with grazing beam of 117 MeV Pb ions at room temperature.

We present in Appendix C an additional analysis of the stoichiometry of the track region in the 10 nm thick film at various irradiation temperatures. The analysis revealed carbonization of the film surfaces, which is consistent with the experimental results [61,62].

### 3.3. Structural changes in SiC films of different thicknesses

In this section, we considered Bi 710 MeV ion irradiation of SiC films with thicknesses ranging from 10 nm to 100 nm across a wide temperature range.

At room irradiation temperature, the track region undergoes the following stages of evolution regardless of the film thickness: (i) formation of a molten cylindrical region, (ii) "squeezing" a part of the melt onto the film surface, and (iii) recrystallization of the molten region. As a result, hillocks were formed on the film surface and nanocavities remained in the film volume along the ion trajectory. However, the shape, size, and crystallinity of the hillocks, as well as the distribution of nanocavities and defect clusters in the film volume change with the film thickness.

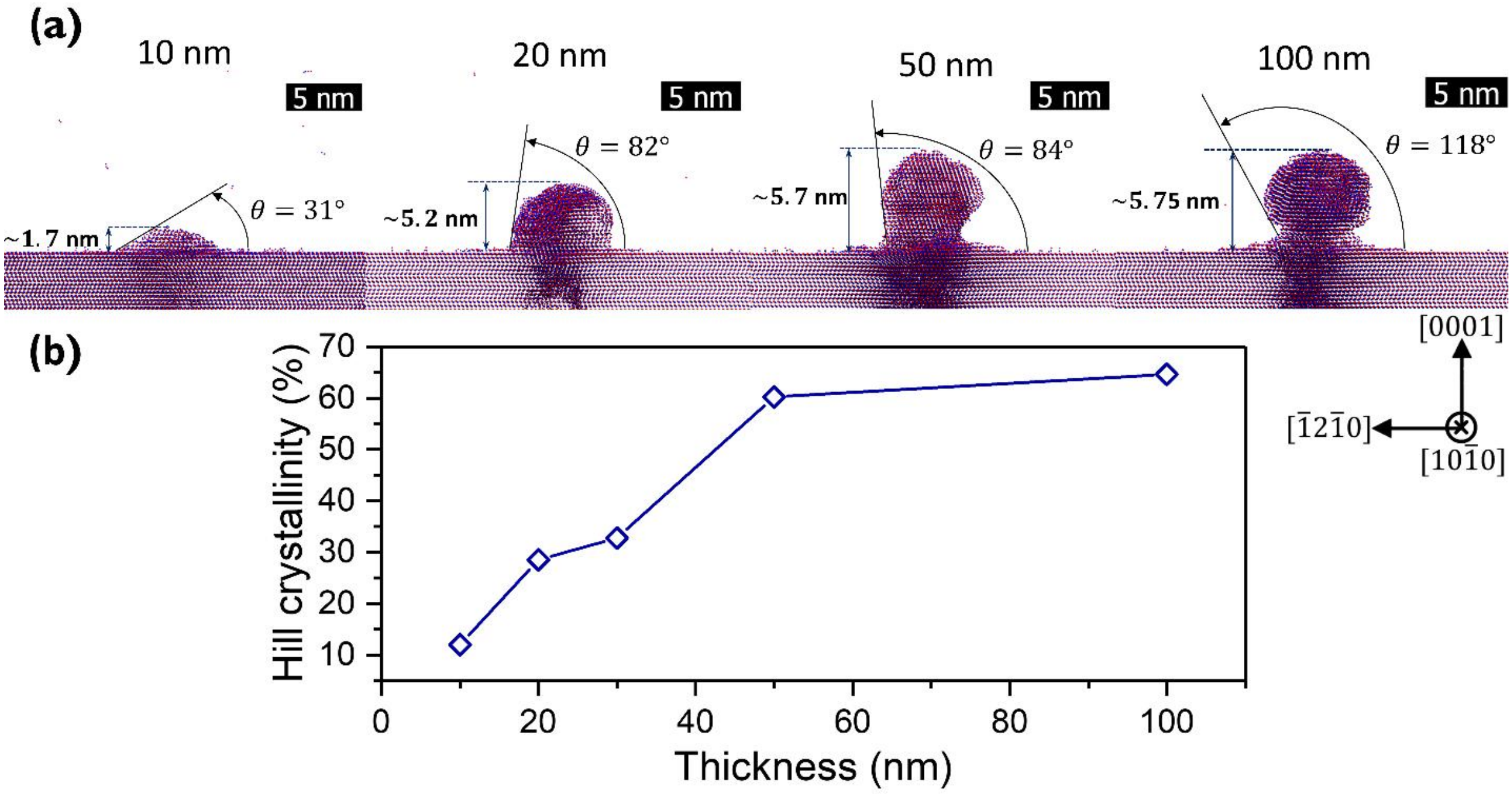

*Fig. 10 (a) Hillocks on the surface of SiC film with thicknesses of 10 nm, 20 nm, 50 nm, and 100 nm after impact of a 710 MeV Bi ion at 300 K, (b) Dependence of hillock crystallinity on the thickness of SiC film irradiated with Bi 710 MeV ion at 300 K.*

Fig. 10a shows that increasing the thickness of the irradiated film from 10 nm to 100 nm leads to an increase in the height of the hillocks on its surface from 1.7 nm to 5.75 nm, as well as an increase in the value of the contact angle from 31° to 118°. The fraction of the crystalline state identical to that in the original SiC crystal film also increases with film thickness from 12% for a 10 nm film to 65% for 100 nm film (see Fig. 10b). Thus, the shape of the hillock and its crystallinity is determined not only by the balance of adhesion and cohesive forces within the melt, as reported in Ref. [63], but also by the amount of molten material that is “squeezed” onto the film surface from the track region. As a result, hillocks of predominantly crystalline composition with a shape close to spherical and a height of about 5.7 nm will be formed on the surface of bulk films after irradiation with 710 MeV Bi ion at room temperature.

Fig. 10 and the conclusions based on it correspond to the C-terminated surface of SiC (upper surface). Similar conclusions also apply to the Si-terminated surface (lower surface), on which nanohillocks with a similar shape and composition but different sizes are formed. The most significant difference was in the 10 nm thick film, where the top nano-hill was 40% larger than the bottom one. However, as the film thickness increased, this difference diminished to 4% in the 100 nm thick film.

The hillock formation on the surface of the films at $T_{irr} = 300$ K is accompanied by structural changes in the deeper regions. In films up to 100 nm thick, a chain of nanocavities surrounded by clusters of defects is formed along the ion trajectory (see Fig. 11). While in 10 nm thick film dislocations are formed in the bulk of the film along the edges of the defect region (see Fig. 7a). In thicker films dislocations were formed only in the 10 nm layer near the surface.

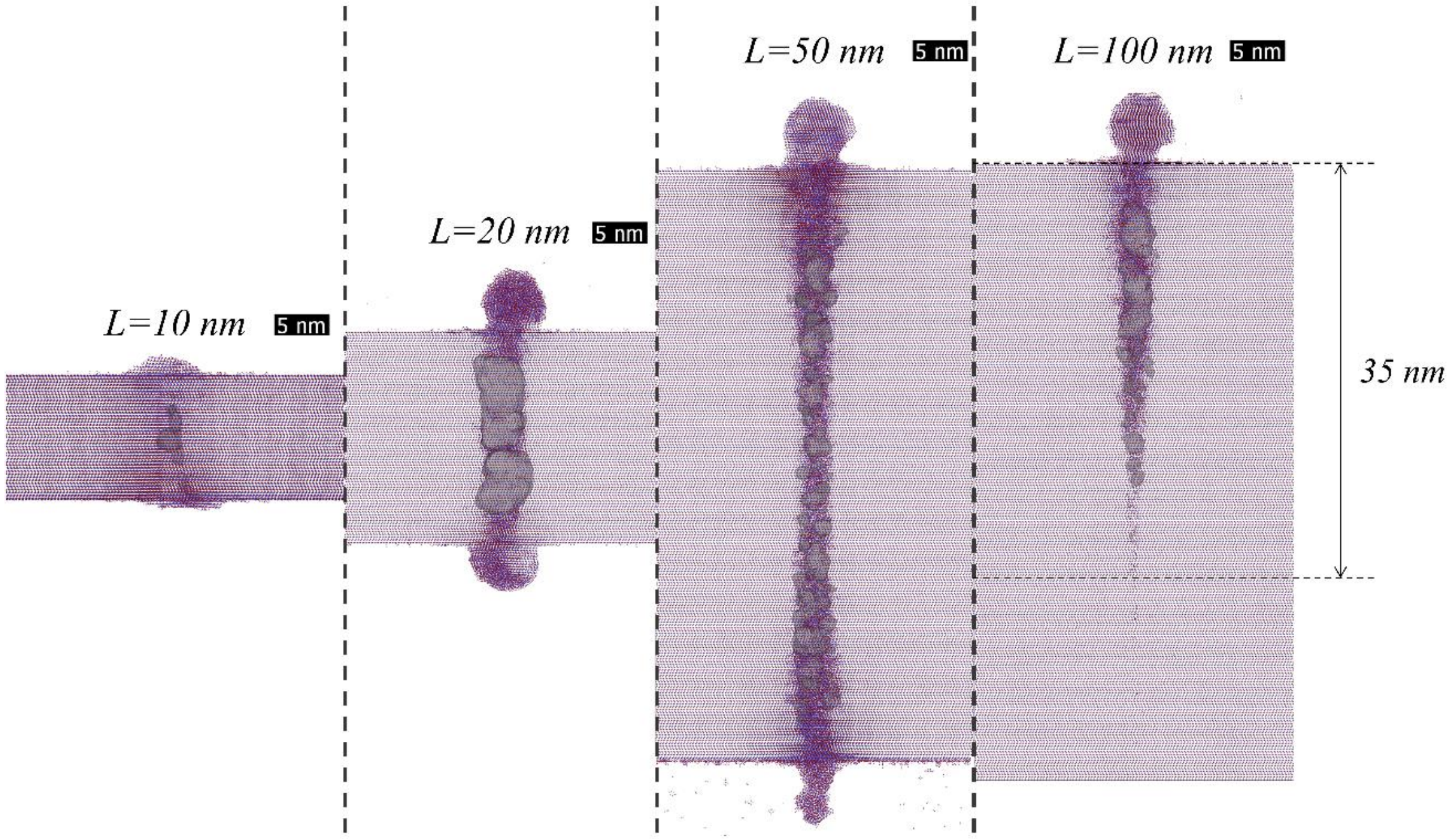

*Fig. 11 Structure changes in SiC films of 10 nm, 20 nm, 50 nm, 100 nm thicknesses with highlighted surfaces of nanocavities after irradiation with 710 MeV Bi ion at room temperature.*

Fig. 11 shows that in a 100 nm thick film the track can be clearly divided into a near-surface region and a bulk region. In the near-surface region, this track consists of nanocavities and defect clusters and has a conical shape with a base diameter of $d \approx 4.4$ nm and a height of $H \approx 35$ nm. In the bulk region, several point defects remain along the ion trajectory, which is in agreement with previous calculations [11,51,55,56] and experimental data [48,64]. The track morphology in such a film at $T_{irr} = 300$ K is very similar to the experimental data from Ref. [62]. This work considers the room-temperature irradiation of SiC fibers with swift heavy ions $^{112}Sn^{27.3+}$ at the energy of 414.4 MeV. The study reports the formation of nanohillocks on the SiC surface, while no structural defects were observed at depths of several micrometers after irradiation. We also observe a qualitative similarity with the experimental results from [65], in which the excitation of the atomic subsystem of SiC was induced by laser irradiation. However, in [65] due to the specificity of laser irradiation, in contrast to the SHI track, nanohillocks are formed around the excited region.

The results of the simulation for the 100 nm film allow to draw qualitative conclusions about the response of multi-layered SiC structures to irradiation and the resulting damage formation. It was believed that SiC-based devices are insensitive to single-event effects due to the resistance of this material to SHI irradiation [48]. However, a number of experimental studies (see Ref. [66] and references therein) have observed structural changes at the metal-dielectric interface in Schottky diodes after SHI irradiation. We demonstrate that SiC melts in the vicinity of the SHI trajectory and a part of the melt is displaced to the surface even at room irradiation temperatures. Next, this melt crystallizes, forming cavities inside the material. In layered nanoscale structures, these cavities themselves can disrupt the electronic structure of the material. In addition, they can be filled with a melt of another substance [67] included in the composition of the diode. Moreover, during the early stages of melt formation (≈5 ps after irradiation in Fig. 6 in our work), metal atoms can form bonds with Si and C atoms [67]. This can impede the effective SiC recovery and lead to an increase in the size of the defect region after solidification, up to the size of the initial melt in the track. Taking all this into account, the track structure formed in the 100 nm film closely resembles the structural modifications observed at the Ti/SiC interface in a Schottky diode irradiated with 1437.6 MeV $Ta^{181}$ ions under a reverse bias voltage of 200 V [66].

In a 100 nm thick film, increasing the irradiation temperature leads to an increase in the size of the defect region with nanocavities and defect clusters deep within the film. Already at 1200 K, this region occupies the entire volume of the film along the ion trajectory.

We also found a transition from hillocks to craters with increasing irradiation temperature for films of different thicknesses. The film thickness increase led to a change from droplet ejection at high irradiation temperatures, which we observed in the 10 nm thick film (see Fig. 9b), to jet ejection, such as shown in Fig. 12 for 100 nm thick film - the thickest of the films we considered.

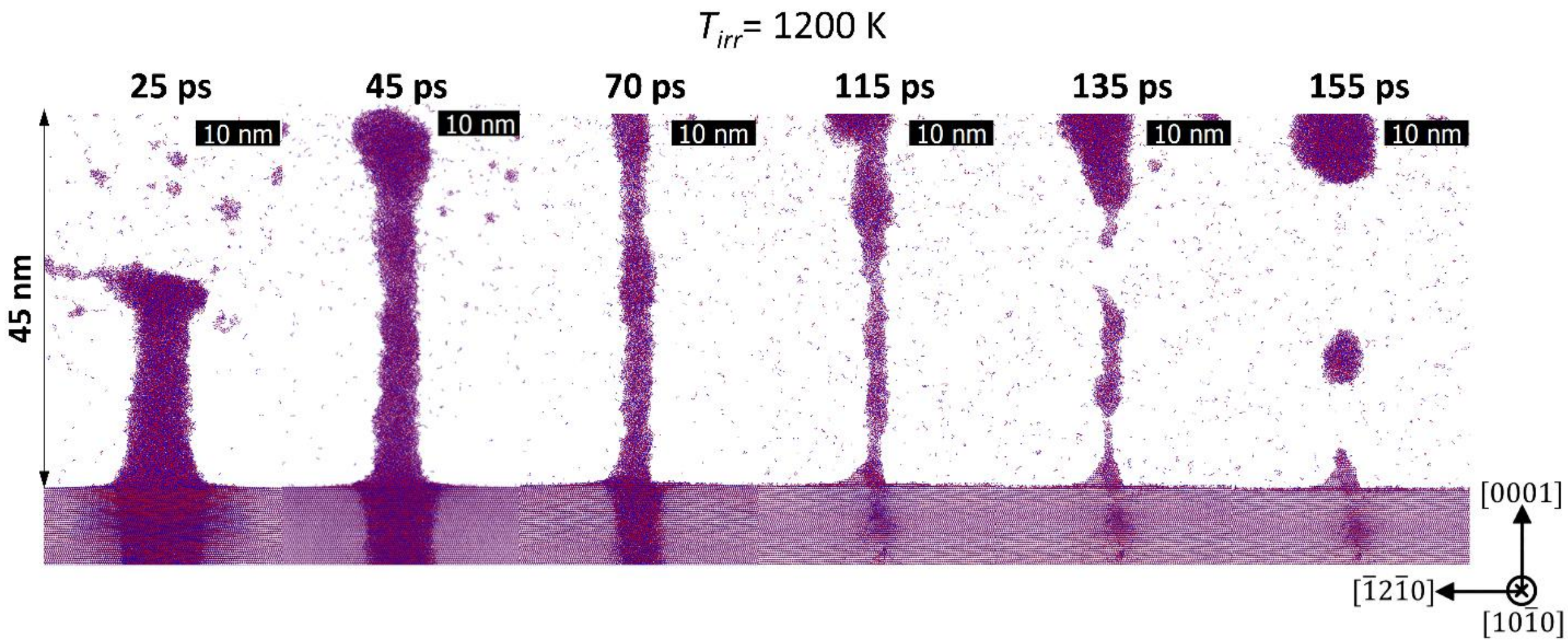

*Fig. 12 Temporal evolution of the near-surface region of a 100 nm thick SiC film after an impact of a 710 MeV Bi ion at 1200 K.*

Fig. 13a shows the calculated dependence of hillock height and crater depth on the irradiation temperature for films of different thicknesses. Based on these data, the transition temperature for each film was determined via linear interpolation between the minimal hillock height and crater depth in the transition region, where "minimal" denotes the values nearest the film surface. Fig. 13b shows the transition irradiation temperature as a function of film thickness. It can be seen that the transition temperature increases with the film thickness.

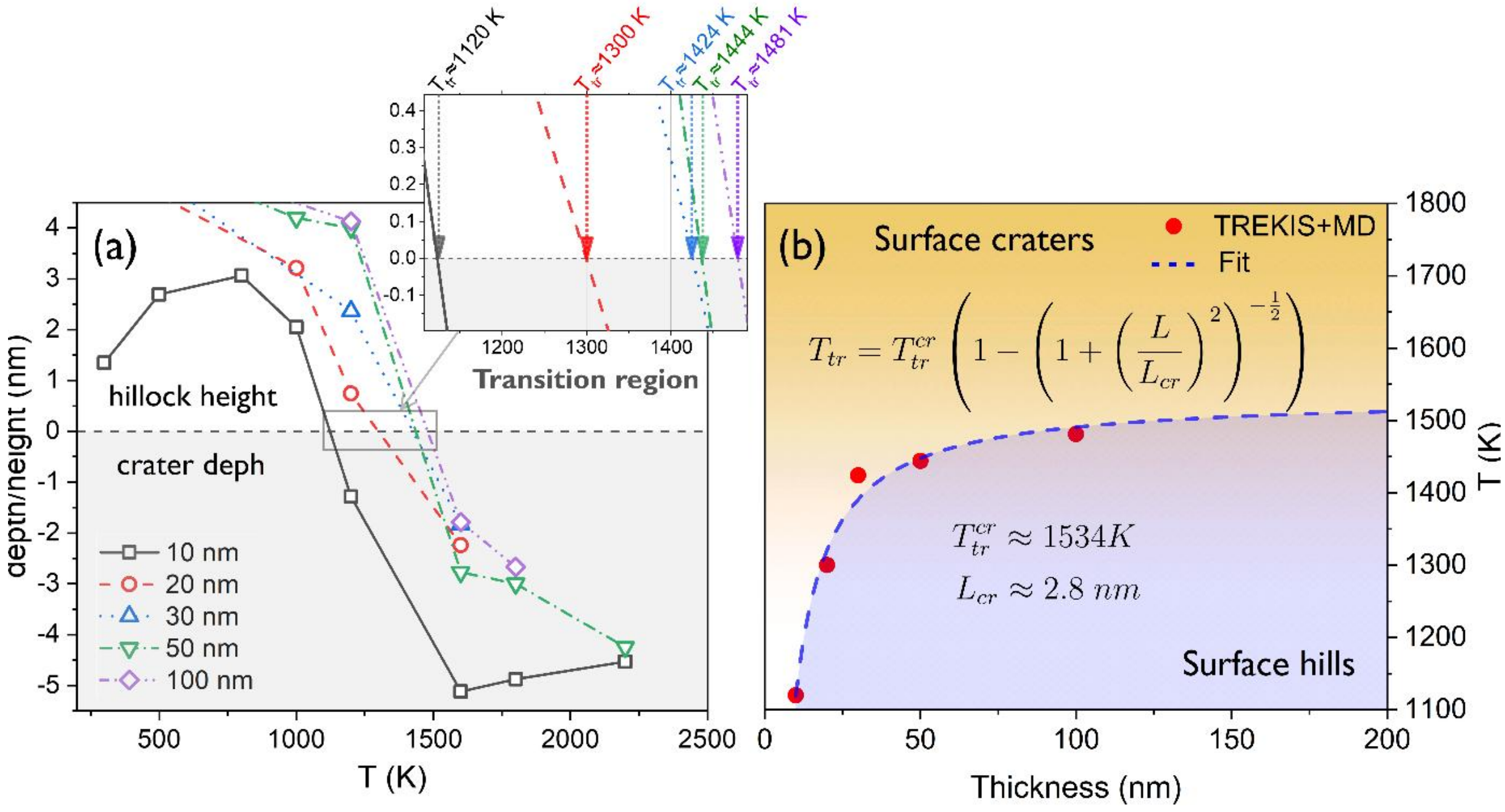

*Fig. 13 (a) Average hillock height (positive) and crater depth (negative) of nanostructures created on the SiC (001) surface after Bi 710 MeV impact at different irradiation temperatures; (b) Transition irradiation temperature as a function of film thickness.*

Apparently, this trend is associated with the threshold amount of substance that must be “ejected” from the film volume in the track region to form craters on its surface. The thicker the film, the stronger the heating of the track region required to suppress resistance of the thicker subsurface layer to ejection of a larger amount of molten material. Thus, the transition temperature should increase with increasing film thickness. We assumed that the threshold lattice excitation in thick films (of about several microns) must be created in a fixed-thickness near-surface narrow layer similar to the dependence of the ablation threshold fluence [68]. Therefore, the threshold energy density and the corresponding transition irradiation temperature should not depend on the thickness of such films.

Consequently, we tested various functions that approach a constant asymptote for thick films and selected the one that best describes our data on the transition temperature as a function of film thickness:

$$T_{tr} = T_{tr}^{cr}\left(1-\left(1+\left(\frac{L}{L_{cr}}\right)^2\right)^{-\frac{1}{2}}\right) \tag{9}$$

Here, $T_{tr}^{cr} \approx 1534$ K, $L_{cr} \approx 2.8$ nm (see Fig. 13b). The quality of our fitting is well illustrated by the coefficient of determination, $R^2 = 0.98$, which is close to unity indicating a very good agreement between Eq. (9) and the calculated transition temperatures. The formula is valid in the range $L > L_{cr}$. Films with thicknesses $L \leq L_{cr}$ were not considered in our work due to the lack of reliable data on their crystalline stability.

The majority of our modeling results fall within the thicknesses $L \gg L_{cr}$, where the transition temperature dependence (9) acquires a nonlinear asymptote: $T_{tr}(L) \approx T_{tr}^{cr}(1 - L_{cr}/L)$. The similar dependence on film thickness was observed for other crystal properties [69,70], for example the melting temperature [71,72].

In our case, this nonlinear asymptotic behavior can be explained by the influence of uncompensated bonds of surface atoms within a layer of thickness on the order of $L_{cr}$ on the film relaxation after irradiation. As the film thickness increases, the fraction of surface atoms decreases as $1/L$, and the transition temperature tends to a constant value $T_{tr}(L \to \infty) = T_{tr}^{cr}$ for bulk targets.

Extrapolation of this expression to the region of films of macroscopic thickness predicts a transition temperature from hillocks to craters of ≈1534 K.

As noted above, the transition temperature for each film was obtained using linear interpolation. Consequently, the choice of a specific interpolation model within the transition region can influence the result. Changing the interpolation model will lead to a uniform shift in transition temperatures, so interpolation affects the scale of the graph in Fig. 13b, but not its shape. Therefore, the error introduced by the interpolation model selection is associated with the parameter $T_{tr}^{cr}$ in Eq. (9), not with the parameter $L_{cr}$, which determines the curve shape. We estimate this error as $T_{tr}^{cr}$ = 1534 ± 100 K.

Thus, function (9) (see Fig. 13b) allowed us to divide the temperature-thickness scale for all films into two regions: (i) the region with hillock formation and (ii) the region with crater formation. However, this separation is approximate, because in films thicker than 30 nm we found the formation of combined structures near the transition temperature: formation of a hillock in the center of the crater on one side of the film, as shown in Fig. 14a, and a crater on the other side. Fig. 14b shows a more detailed height map, which allows for a more precise determination of the hillock position relative to the crater.

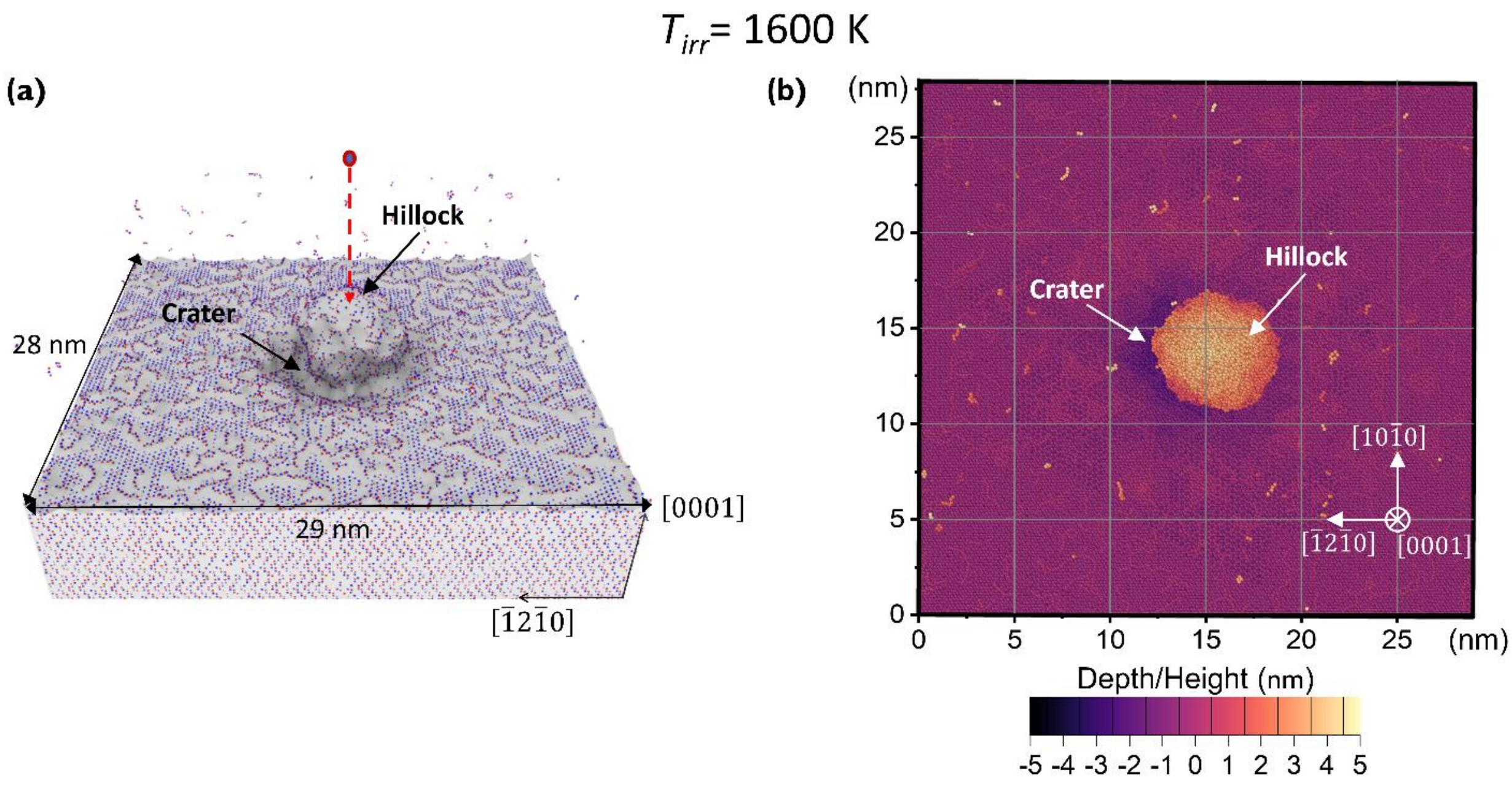

*Fig. 14 (a) Hillock formation inside a crater in a 100 nm thick SiC film after an impact of a 710 MeV Bi ion at 1600 K (b) Height map of this region.*

At the beginning of this section, we have demonstrated that the formation of nanohillocks at room irradiation temperature is caused by high pressure in the hot ion track. This pressure is associated with the low density of the molten SiC compared to the crystalline one independently of its specific structure. Taking this into account, we suggest that nanohillocks should also form in other SiC polytypes during room-temperature irradiation, provided they have almost the same ratio between the densities of the melt and crystalline phases. At high irradiation temperatures, we showed that nanocraters are formed via the ejection of material droplets or jets from the target, depending on its thickness. We observed no influence of the crystal symmetry on this ejection. Again, we conclude that there is a high probability that craters should also be observed in other polytypes at high irradiation temperatures. Therefore, we believe that the transition from the nanohillocks to craters with irradiation temperature is highly likely to occur in silicon carbide polytypes other than 6H-SiC.

On the other hand, structural changes in SiC films are determined not only by interatomic interactions, but also by the spatial profile of the energy transferred to the lattice in the SHI track. This profile, in turn, depends on the spectrum of electrons excited by SHI and their scattering on the atomic system (the rate of electron-lattice coupling). As a result, the spatial distribution of energy deposited in the atomic lattice is determined by the phonon spectrum in SiC, the structure of the valence band, the work function, and their temperature dependencies. All these parameters differ among the SiC polytypes [73]. This means, that quantitative analysis of the shape and size of nanohillocks at room irradiation temperature, as well as determination of the transition temperature from the hillocks to craters, requires separate consideration for each polytype.

## 4. Conclusions

We observed a transition from nanohillocks to nanocraters on the surface of SiC films when simulating impacts of 710 MeV Bi ions. This transition occurs as the irradiation temperature increases. For films with thicknesses ranging from 10 nm to 100 nm, we determined the irradiation temperatures at which this transition occurs and established the dependence of the transition irradiation temperature on film thickness. The transition temperature increases with the film thickness from ≈1120 K for a 10 nm thick film to ≈1481 K for a 100 nm film. Extrapolation this dependence predicts a transition irradiation temperature of ≈1534± 100 K for the bulk SiC.

Our results can be used for controlled SiC surface texturing. The formation of nanohillocks on the surface of SiC may reduce its reflectance [74], that can be used in the creation of energy conversion devices. Following Ref. [5,75], the creation of surface craters may increase the light extraction efficiency in fluorescent SiC based all semiconductor LED light sources [76].

The results can also be used for planning experiments on SiC film irradiations with SHIs at temperatures higher than 1000 K. Such experiments will provide both: detail information about damage of the films at such condition as well as verification of the simulation results.

On the other hand, we think that the formation of craters and hillocks during irradiation process (material removal) should lead to a significant decrease in the film's thermal conductivity. This effect is probably associated with the existence of nanocavities and defective structures in the near-surface region, which hinders phonon heat transfer [58]. This, in turn, may lead to premature failure of SiC-based electronics under radiation conditions.

## Code and data availability

The free TREKIS-3 [23,24] code, used to model the excitation of the electronic and ionic systems in the SHI track, is available from [77]. The LAMMPS code was used for molecular dynamic calculations [26]. Ovito [78] software was used for visualization of MD calculations.

## Acknowledgements

This work has been carried out using computing resources of the federal collective usage center Complex for Simulation and Data Processing for Mega-science Facilities at NRC "Kurchatov Institute", http://ckp.nrcki.ru/. We are grateful to S.A. Gorbunov for helpful discussions and comments on the paper.

## Appendix A

This section presents an analysis of the effect of temperature dependence of the work function on structural modifications in SiC films under high-temperature irradiation (1500–2200 K). Ref. [45] demonstrated that the work function of silicon carbide weakly depends on the crystal structure of SiC at high temperatures. Therefore, to estimate the temperature dependence of the work function, we can use experimental data from Ref. [44], obtained in the temperature range of 1500–2100 K for highly dispersed SiC powder: $W(T) = 4.37 + 1.2 \times 10^{-4} T$(K). Note, that at 1500 K, this formula predicts a work function of 4.55 eV, which is practically identical to the work function of hexagonal 6H-SiC at this temperature. Further extrapolation leads to a ~2% increase in the work function with the temperature increase from 300 K to 2200 K. Additional analysis performed for a 10 nm thick film showed that such an increase does not alter the number of emitted electrons (approximately 3.9 thousand). However, the profile of the energy deposited into the lattice (Fig. A. 1) shifted toward the profile calculated with periodic boundary conditions which are equivalent to an infinite work function excluding electron emission. In this case, the energy density deposited in the lattice within the central region of the track is only 8% greater than for the barrier with the work function of 4.5 eV.

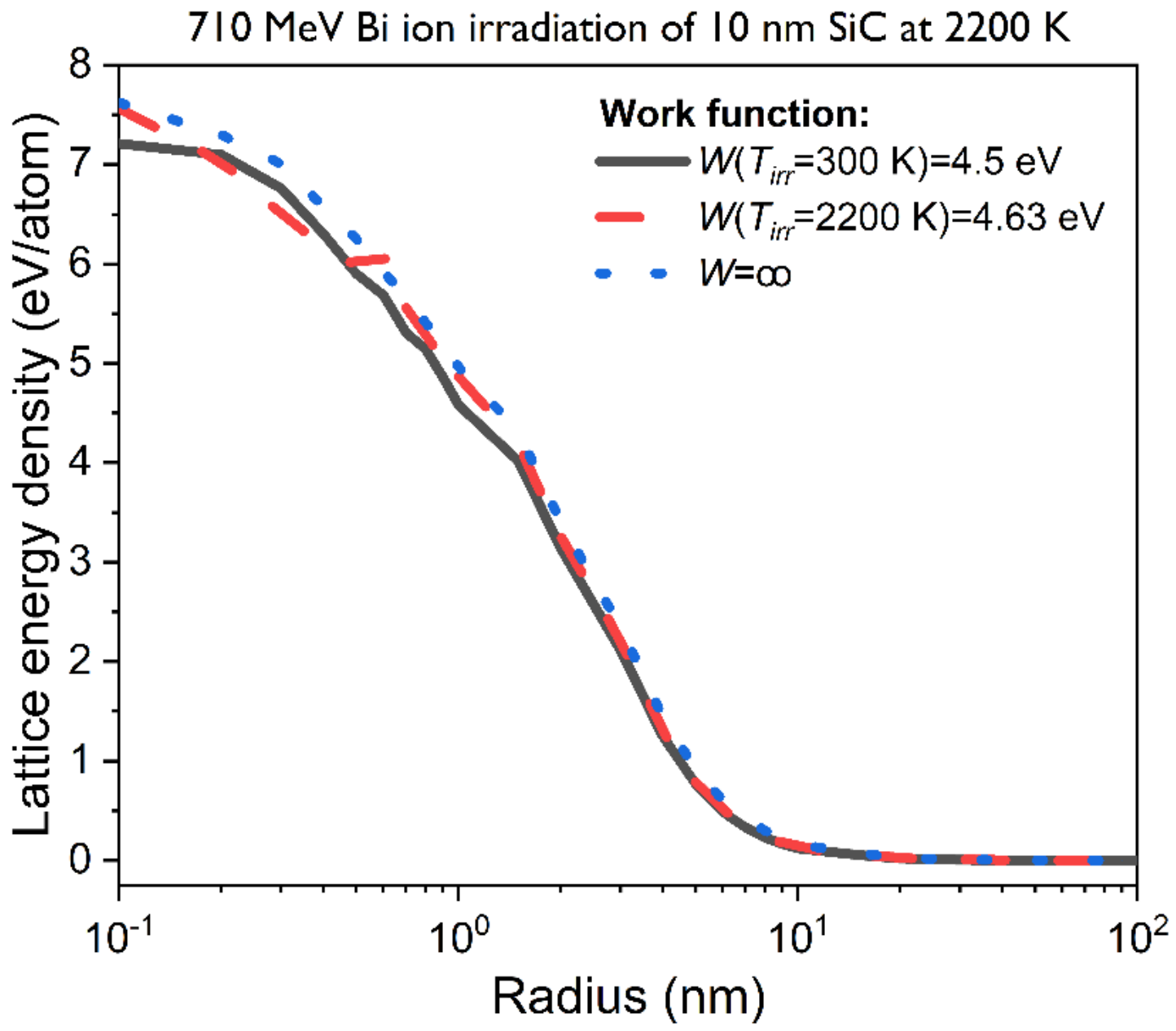

*Fig. A. 1 Profile of the energy deposited to atomic lattice of 10 nm thick SiC film after 710 MeV Bi impact at 2200 K for different work function values.*

To assess how the temperature dependence of the work function can affect structural changes in the track, we performed molecular dynamics simulations of irradiation of a 10 nm thick film at 2200 K for two limiting cases: (i) with electron emission using a temperature-independent work function ($W$ = 4.5 eV), and (ii) without electron emission, corresponding to the infinite work function. Fig. A. 2 shows that the crater depth of 4.49 nm in the track, obtained with electron emission and $W$=4.5 eV, almost coincides with the depth of 4.53 nm obtained without electron emission. The difference in depth is negligible, being less than 1%. Consequently, electron emission, and therefore the temperature dependence of the barrier, does not affect the crater depth. Since our study focuses only on crater depths, electron emission does not affect the conclusions of our work at high temperatures (>1500 K). On the other hand, Fig. A. 2 shows that accounting for electron emission with $W$ = 4.5 eV affects the shape of craters by making their edges steeper. Consequently, for a more accurate description of the crater shape, rather than just its depth, it is necessary to consider electron emission in the track as well as the temperature dependence of the work function. However, this lies beyond the scope of the present study.

Since the precise temperature dependence of the work function in 6H-SiC is unknown, we use a barrier with a work function of 4.5 eV in all simulations that account for electron emission. Moreover, the influence of electron emission diminishes with increasing film thickness due to the decreasing relative contribution of the near-surface layer compared to the bulk. As a result, we conclude that using the infinite barrier approximation excluding electron emission for films thicker than 30 nm is also justified.

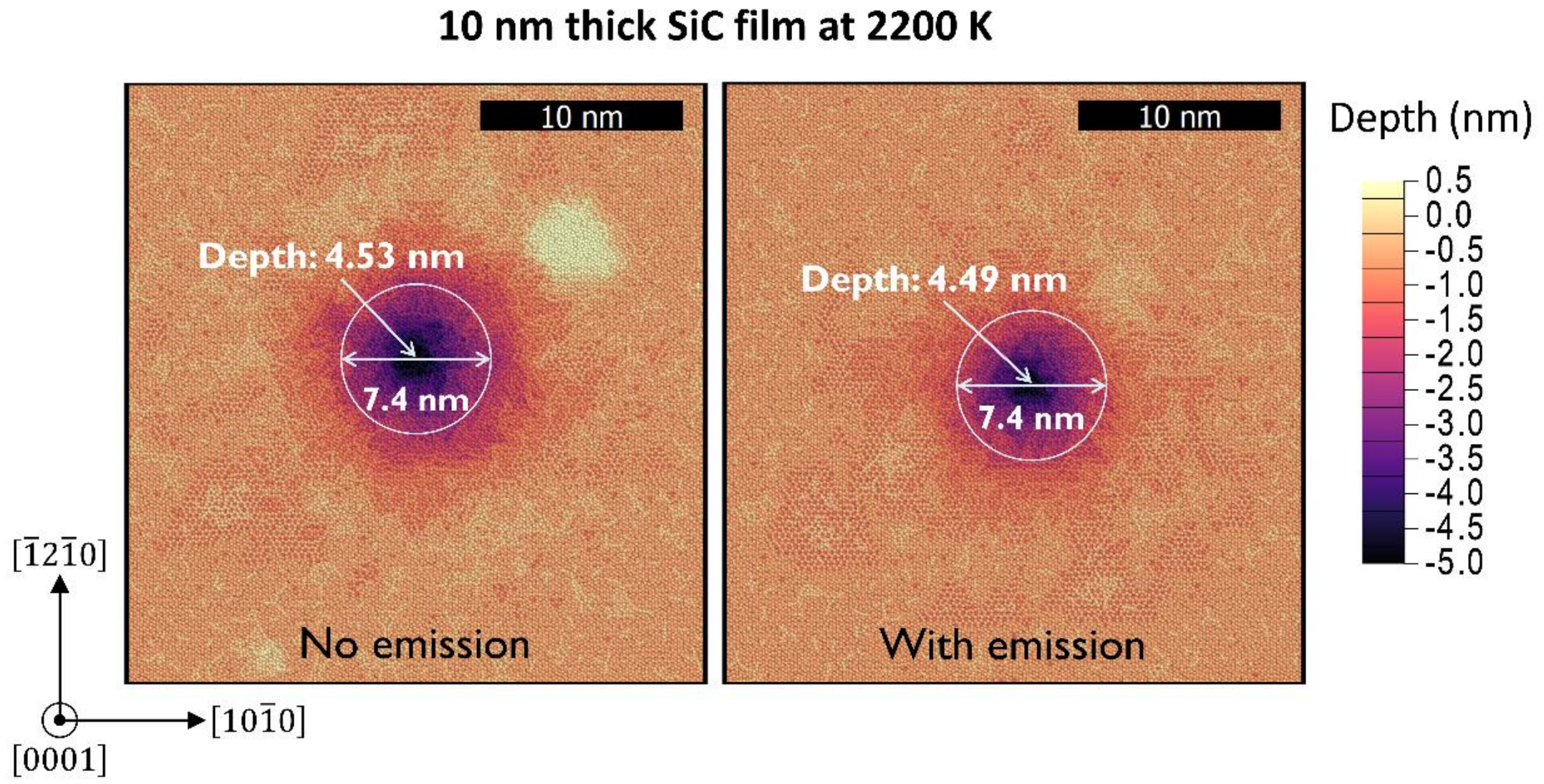

*Fig. A. 2 Height maps of a 10 nm thick SiC film irradiated with a 710 MeV Bi ion at 2200 K: comparison of simulations made with electron emission (W=4.5 eV) and without it.*

# Appendix B

In this section, we present an additional analysis of how changes in ion energy and type could affect the boundaries of the four irradiation regimes and the transition temperature described in Section 3.2.

The ejection of material droplets from the sample surface leading to the transition from hillocks to craters, is initiated by a threshold degree of excitation of the material in the track region. It depends on the energy transferred from the excited electrons to the target atoms. This energy is determined by two factors: (a) the parameters of the incident ion (charge, velocity, mass, incident angle) and (b) the irradiation temperature, which determines the rate of electron-phonon energy exchange as well as indicates the “starting” state for the material excitation in an ion track. As a result, the threshold irradiation temperatures that stimulate droplet ejection can differ for different ions.

In addition to 710 MeV Bi we performed simulations of the irradiation of a 10 nm thick film with a 167 MeV Xe ion, which is also widely used in experiments. The 65% difference in electronic energy loss between the 167 MeV Xe ion ($dE_e/dx$ ≈21.2 keV/nm) and the 710 MeV Bi ion ($dE_e/dx$ ≈35 keV/nm) led to an increase in the transition temperature by only ≈1.3%, from 1120 K for 710 MeV Bi to 1134 K for 167 MeV Xe (see Fig. B. 1). This shows a weak dependence of the transition temperature on energy losses. In other words, the transition temperature depends not only on energy loss but, e.g. also on the spectrum of excited electrons. This is similar to the velocity effect [21], where lower energy loss does not always result in less damage.

However, as seen in Fig. B. 1, the four temperature regimes of irradiation described in our work for Bi 710 MeV differ significantly from the corresponding regimes for Xe 167 MeV. Moreover, due to lower energy losses for Xe compared to Bi, these regimes are less pronounced. Further investigation of the sensitivity of the transition temperature to the ion charge and its energy, in particular its connection with the velocity effect [21], is beyond the scope of this work and is a topic for a separate study.

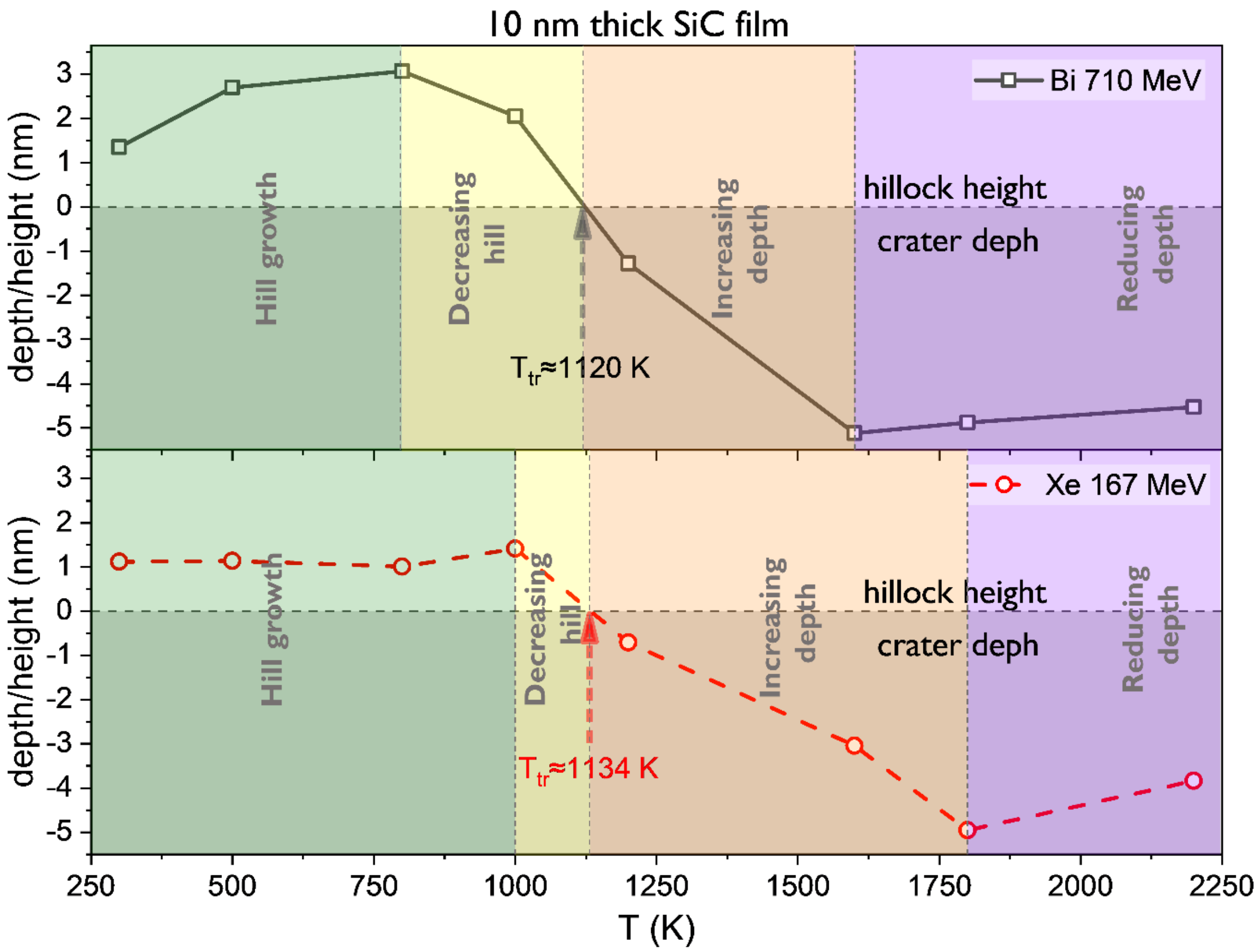

*Fig. B. 1 Height of hills and depth of craters (relative to the film surface, where zero corresponds to the surface) as a function of irradiation temperature for a 10 nm thick film irradiated with 710 MeV Bi and 167 MeV Xe ions. The four irradiation regimes are color-coded: (i) hill height increases with temperature (Hill growth), (ii) hill height decreases with temperature (Decreasing hill), (iii) crater depth increases (Increasing depth), (iv) crater depth decreases (Reducing depth).*

# Appendix C

In this section, we investigate the temporal evolution of stoichiometry in the track region for a 10 nm thick film irradiated by 710 MeV Bi ions at four different temperatures (see Fig. 8), corresponding to different irradiation regimes described in Section 3.2.

We investigated the composition of the molten region in SHI tracks and the composition of the defective region remaining after the track solidification. Since these regions are associated with local symmetry breaking, we identified atoms belonging to these regions using the centrosymmetry parameter (CSP) [79], which was calculated for each atom separately using the formula:

$$CSP = \sum_{i=1}^{\frac{N}{2}} \left| \boldsymbol{r}_i + \boldsymbol{r}_{i+\frac{N}{2}} \right|^2 \qquad (C.1)$$

where $\boldsymbol{r}_i$ and $\boldsymbol{r}_{i+\frac{N}{2}}$ are vectors from the central atom to neighbor atoms, to the *i*-th atom and its opposite atom, respectively, and the sum runs over *N/2* nearest neighbors. In the 6H-SiC, *N*=4 because each atom in the lattice is tetrahedrally bonded to four atoms of another type [80]. Since there is no inversion symmetry in a tetrahedral system, atoms in defect regions and in the melt will have a lower CSP value than atoms in the virgin crystal lattice. To analyze defect clusters in the track core, we selected atoms with CSP<5 $Å^2$. This condition, which depends on the distance to the four nearest neighbors, allowed us to exclude atoms of the ideal crystal lattice, as well as atoms for which CSP was disrupted by thermal vibrations of atoms or isolated point defects. Atoms ejected from the film were also excluded from consideration. For the analyzed atoms, we did not observe any significant difference in stoichiometry from the virgin lattice. A slight predominance of Si, approximately 2% more than C, in the initial melt in the track was associated with the incongruent melting of SiC [81], in which the Si fraction turns into a liquid and the C fraction remains in a solid.

We also monitored the stoichiometry of atoms on the film surface and on the surfaces of cavities formed in the SiC films after irradiation. To identify these surface atoms, we considered their coordination number, which is smaller than that of atoms in the crystal. In the film, we considered atoms for which the number of neighbors within a distance of 3.5 Å, taking into account the next nearest neighbors (NNN), was less than 12 but greater than 5. This excludes single atoms or clusters of atoms that were ejected from the film during irradiation. Larger clusters ejected from the film were also excluded from consideration. Fig. C. 1a shows the selected atoms on the surfaces of the 10 nm thick film 200 ps after irradiation with a 710 MeV Bi ion at four different temperatures. Among the atoms we identified in the early stages of track formation there were atoms included in the initially molten region, since the density of this region was lower than that of the crystalline region. Therefore, due to the incongruent melting of SiC [81], as noted earlier, the proportion of silicon in the initial melt in the track exceeded the proportion of carbon by approximately 2%, regardless of the irradiation temperature (see Fig. C. 1b).

Simulations of the room-temperature irradiation of a 10 nm thick film, Fig. C. 1 shows that defective structures with a high carbon fraction were formed on the film surfaces from a melt that was approximately homogeneous in composition. At the same time, the change in the carbon fraction with time occurred in a similar manner on both the outer surfaces of the film and the inner surfaces of the cavities. To increase the statistical reliability of the analysis of surface carbonization, both of these regions were taken into account.

No material ejection from the film was observed during the simulation at room irradiation temperature, and we think that the carbonization of the surfaces is not associated with the ejection of silicon. This conclusion is consistent with the experimental data [82] in which carbonization of the 6H-SiC surface occurred at a high annealing temperature without material evaporation.

The surface carbonization observed in the simulation is also consistent with the results of experiments [61,62], which also show an increase in carbon concentration on the surface after irradiation at room temperature.

At high irradiation temperatures, modeling also showed an increase in the carbon concentration on the surfaces, but it was significantly smaller than that at room irradiation temperature. However, among the atoms ejected from the film, Si atoms were only 1% more than C atoms, and the carbonization of the surfaces in the ion tracks cannot be explained by a deficiency of Si in the film.

We associate the observed equalization of the composition on surfaces at high irradiation temperatures with an increase of the entropic term in the free energy of the system [83]. At room irradiation temperature, the role of this term decreases as the track cools down. This brings the system into the temperature range optimal for carbonization [82]. As a result, a kink appears in Fig. C. 1b.

Our calculations may underestimate the carbon fraction in the track region on SiC surfaces because we do not take into account the kinetics of defects at long times after irradiation (>200 ps). In addition, the interatomic potential we use perfectly describes the reaction of SiC to extreme excitation [11], but it is not designed to describe the subtle effects associated with the migration of point defects on large time scales. Special potentials must be used for these tasks. For example, the use of such a special potential [84] shows that the migration of carbon interstitials over long times can affect the stoichiometry of final defects.

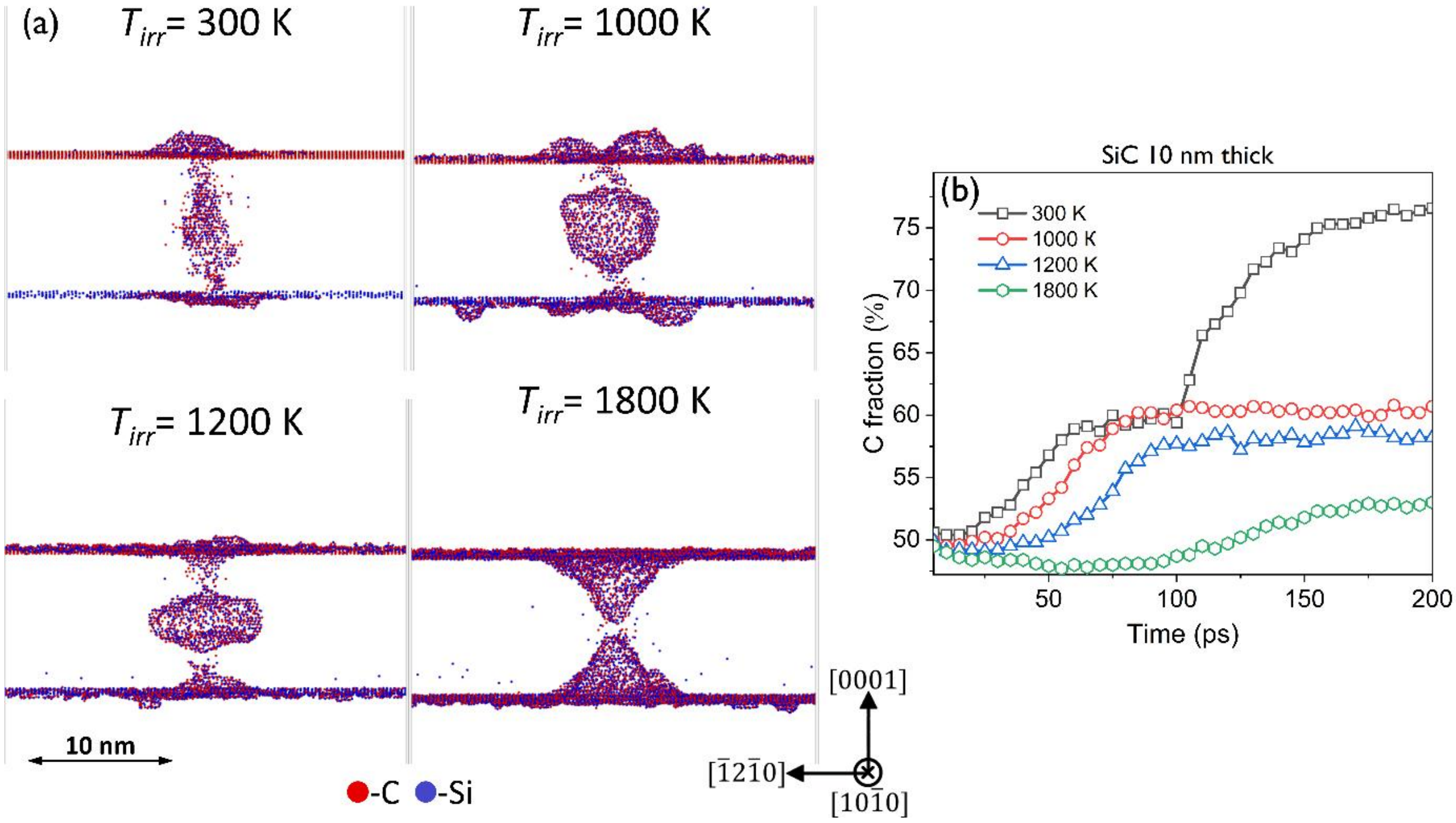

*Fig. C. 1 Carbon fraction on the surfaces of a 10 nm thick SiC film after impacts of 710 MeV Bi ion at various temperatures.*

## Author contributions: CRediT

**D.I. Zainutdinov:** Conceptualization; Formal Analysis; Methodology; Visualization; Writing – original draft; Writing – review & editing. **A.E. Volkov:** Conceptualization; Formal Analysis; Methodology; Project administration; Writing – review & editing.

## References

[1] A.E. Kaloyeros, B. Arkles, Silicon Carbide Thin Film Technologies: Recent Advances in Processing, Properties, and Applications - Part I Thermal and Plasma CVD, ECS J. Solid State Sci. Technol. 12 (2023) 103001. https://doi.org/10.1149/2162-8777/acf8f5.

[2] F.H. Ruddy, L. Ottaviani, A. Lyoussi, C. Destouches, O. Palais, C. Reynard-Carette, Silicon Carbide Neutron Detectors for Harsh Nuclear Environments: A Review of the State of the Art, IEEE Trans. Nucl. Sci. 69 (2022) 792–803. https://doi.org/10.1109/TNS.2022.3144125.

[3] X. Zhang, Y. Cao, C. Chen, L. Wu, Z. Wang, S. Su, W. Zhang, L. Lv, X. Zheng, W. Tian,

X. Ma, Y. Hao, Study on Single Event Effects of Enhanced GaN HEMT Devices under Various Conditions, Micromachines 15 (2024). https://doi.org/10.3390/mi15080950.

[4] M. De Napoli, SiC detectors: A review on the use of silicon carbide as radiation detection material, Front. Phys. 10 (2022) 1–28. https://doi.org/10.3389/fphy.2022.898833.

[5] F. La Via, D. Alquier, F. Giannazzo, T. Kimoto, P. Neudeck, H. Ou, A. Roncaglia, S.E. Saddow, S. Tudisco, Emerging SiC Applications beyond Power Electronic Devices, Micromachines (2023) 1–37.

[6] A.F. Witulski, D.R. Ball, K.F. Galloway, A. Javanainen, J.-M. Lauenstein, A.L. Sternberg, R.D. Schrimpf, Single-Event Burnout Mechanisms in SiC Power MOSFETs, IEEE Trans. Nucl. Sci. 65 (2018) 1951–1955. https://doi.org/10.1109/TNS.2018.2849405.

[7] C. Liu, G. Guo, H. Shi, Z. Zhang, F. Li, Y. Zhang, J. Han, Study on the Single-Event Burnout Effect Mechanism of SiC MOSFETs Induced by Heavy Ions, Electronics 13 (2024) 3402. https://doi.org/10.3390/electronics13173402.

[8] Y. Katoh, L.L. Snead, Silicon carbide and its composites for nuclear applications – Historical overview, J. Nucl. Mater. 526 (2019) 151849. https://doi.org/10.1016/j.jnucmat.2019.151849.

[9] A. Debelle, L. Thomé, I. Monnet, F. Garrido, O.H. Pakarinen, W.J. Weber, Ionization-induced thermally activated defect-annealing process in SiC, Phys. Rev. Mater. 3 (2019) 1–11. https://doi.org/10.1103/PhysRevMaterials.3.063609.

[10] E. Kucal, P. Jóźwik, C. Mieszczyński, R. Heller, S. Akhmadaliev, C. Dufour, K. Czerski, Temperature Effects of Nuclear and Electronic Stopping Power on Si and C Radiation Damage in 3C-SiC, Materials (Basel). 17 (2024) 2843. https://doi.org/10.3390/ma17122843.

[11] D.I. Zainutdinov, V.A. Borodin, S.A. Gorbunov, N. Medvedev, R.A. Rymzhanov, M.V. Sorokin, R.A. Voronkov, A.E. Volkov, High-temperature threshold of damage of SiC by swift heavy ions, J. Alloys Compd. 1013 (2024) 1–29. https://doi.org/10.1016/j.jallcom.2025.178524.

[12] D.I. Zainutdinov, R.A. Voronkov, S.A. Gorbunov, N. Medvedev, R.A. Rymzhanov, M. V. Sorokin, A.E. Volkov, Modeling of Temperature Effects on the Formation of Tracks of Swift Heavy Ions in Silicon Carbide, J. Surf. Investig. X-Ray, Synchrotron Neutron Tech. 2024 183 18 (2024) 683–689. https://doi.org/10.1134/S1027451024700319.

[13] R.A. Rymzhanov, N. Medvedev, A.E. Volkov, Damage kinetics induced by swift heavy ion impacts onto films of different thicknesses, Appl. Surf. Sci. 566 (2021) 150640. https://doi.org/10.1016/j.apsusc.2021.150640.

[14] R.A. Rymzhanov, A.E. Volkov, V.A. Skuratov, Bulk, overlap and surface effects of swift heavy ions in CeO2, J. Nucl. Mater. 604 (2025) 155480.

[15] N. Ishikawa, N. Okubo, T. Taguchi, Experimental evidence of crystalline hillocks created by irradiation of CeO2 with swift heavy ions: TEM study, Nanotechnology 26 (2015) 355701. https://doi.org/10.1088/0957-4484/26/35/355701.

[16] D.I. Zainutdinov, A.E. Volkov, Effect of the SiC Film Thickness on the Morphology of Swift Heavy Ion Tracks, Crystallogr. Reports 70 (2025) 1194–1200. https://doi.org/10.1134/S1063774525601133.

[17] Z.G. Wang, C. Dufour, E. Paumier, M. Toulemonde, The Se sensitivity of metals under swift-heavy-ion irradiation: A transient thermal process, J. Phys. Condens. Matter 6 (1994) 6733–6750. https://doi.org/10.1088/0953-8984/6/34/006.

[18] D.M. Duffy, A.M. Rutherford, Including the effects of electronic stopping and electron-ion interactions in radiation damage simulations, J. Phys. Condens. Matter 19 (2007) 0–11. https://doi.org/10.1088/0953-8984/19/1/016207.

[19] R. Darkins, P.W. Ma, S.T. Murphy, D.M. Duffy, Simulating electronically driven structural changes in silicon with two-temperature molecular dynamics, Phys. Rev. B 98 (2018) 1–11. https://doi.org/10.1103/PhysRevB.98.024304.

[20] R.L. Fleischer, P.B. Price, R.M. Walker, Ion Explosion Spike Mechanism for Formation of Charged-Particle Tracks in Solids, J. Appl. Phys. 36 (1965) 3645. https://doi.org/10.1063/1.1703059.

[21] N. Medvedev, A.E. Volkov, R. Rymzhanov, F. Akhmetov, S. Gorbunov, R. Voronkov, P. Babaev, Frontiers, challenges, and solutions in modeling of swift heavy ion effects in materials, J. Appl. Phys. 133 (2023) 100701. https://doi.org/10.1063/5.0128774.

[22] G. Schiwietz, K. Czerski, M. Roth, F. Staufenbiel, P.L. Grande, Femtosecond dynamics – snapshots of the early ion-track evolution, Nucl. Instruments Methods Phys. Res. Sect. B Beam Interact. with Mater. Atoms 226 (2004) 683–704. https://doi.org/10.1016/j.nimb.2004.05.043.

[23] N.A. Medvedev, R.A. Rymzhanov, A.E. Volkov, Time-resolved electron kinetics in swift heavy ion irradiated solids, J. Phys. D. Appl. Phys. 48 (2015) 355303. https://doi.org/10.1088/0022-3727/48/35/355303.

[24] R.A. Rymzhanov, N.A. Medvedev, A.E. Volkov, Effects of model approximations for electron, hole, and photon transport in swift heavy ion tracks, Nucl. Instruments Methods Phys. Res. Sect. B Beam Interact. with Mater. Atoms 388 (2016) 41–52. https://doi.org/10.1016/j.nimb.2016.11.002.

[25] R.A. Rymzhanov, N.A. Medvedev, A.E. Volkov, Electron emission from silicon and germanium after swift heavy ion impact, Phys. Status Solidi B 252 (2015) 159–164. https://doi.org/10.1002/pssb.201400130.

[26] S. Plimpton, Fast Parallel Algorithms for Short-Range Molecular Dynamics, J. Comput. Phys. 117 (1995) 1–19. https://doi.org/10.1006/jcph.1995.1039.

[27] F. Mercier, S. Nishizawa, Role of surface effects on silicon carbide polytype stability, J. Cryst. Growth 360 (2012) 189–192. https://doi.org/10.1016/j.jcrysgro.2011.11.052.

[28] Z. Liu, X. Cheng, X. Yang, M. Liu, R. Liu, B. Liu, Y. Tang, Ultra-high temperature microstructural changes of SiC layers in TRISO particles, Ceram. Int. 50 (2024) 2331–2339. https://doi.org/https://doi.org/10.1016/j.ceramint.2023.11.007.

[29] N.A. Medvedev, A.E. Volkov, Nonthermal acceleration of atoms as a mechanism of fast lattice heating in ion tracks, J. Appl. Phys. 225903 (2022) 131. https://doi.org/10.1063/5.0095724.

[30] L. Van Hove, Correlations in Space and Time and Born Approximation Scattering in Systems of Interacting Particles, Phys. Rev. 95 (1954) 249–262. https://doi.org/10.1103/PhysRev.95.249.

[31] L. Johansson, F. Owman, P. Mårtensson, C. Persson, U. Lindefelt, Electronic structure of 6H-SiC(0001), Phys. Rev. B - Condens. Matter Mater. Phys. 53 (1996) 13803–13807. https://doi.org/10.1103/PhysRevB.53.13803.

[32] A. Chvála, R. Szobolovszký, J. Kovác, M. Florovic, J. Marek, L. Cernaj, D. Donoval, C. Dua, S.L. Delage, J.C. Jacquet, Advanced characterization techniques and analysis of thermal properties of AlGaN/GaN Multifinger Power HEMTs on SiC substrate supported by three-dimensional simulation, J. Electron. Packag. Trans. ASME 141 (2019).

https://doi.org/10.1115/1.4043477.

[33] M. Bauer, A.M. Gigler, A.J. Huber, R. Hillenbrand, R.W. Stark, Temperature-depending Raman line-shift of silicon carbide, J. Raman Spectrosc. 40 (2009) 1867–1874. https://doi.org/10.1002/jrs.2334.

[34] P.H. Berens, D.H.J. Mackay, G.M. White, K.R. Wilson, Thermodynamics and quantum corrections from molecular dynamics for liquid water, J. Chem. Phys. 79 (1983) 2375–2389. https://doi.org/10.1063/1.446044.

[35] J.S. Bader, B.J. Berne, Quantum and classical relaxation rates from classical simulations, J. Chem. Phys. 100 (1994) 8359–8366. https://doi.org/10.1063/1.466780.

[36] L. Frommhold, COLLISION-INDUCED ABSORPTION IN GASES, 1993.

[37] C. Kittel, C.Y. (Ching-yao) Fong, Quantum theory of solids, Wiley, 1987.

[38] W.H. Barkas, Nuclear research emulsions., Academic Press, New York, 1963. http://hdl.handle.net/2027/uc1.b3533401 (accessed November 20, 2013).

[39] C. Bouchard, J.D. Carette, The Surface Potential Barrier in Secondary Emission from Semiconductors, Surf. Sci. 100 (1980) 251–268.

[40] C. Eckart, The penetration of a potential barrier by electrons, Phys. Rev. 35 (1930) 1303–1309. https://doi.org/10.1103/PhysRev.35.1303.

[41] R.A. Rymzhanov, M. Ćosić, N. Medvedev, A.E. Volkov, From groove to hillocks – Atomic-scale simulations of swift heavy ion grazing impacts on CaF2, Appl. Surf. Sci. 652 (2024). https://doi.org/10.1016/j.apsusc.2024.159310.

[42] H. Vázquez, A. Kononov, A. Kyritsakis, N. Medvedev, A. Schleife, F. Djurabekova, Electron cascades and secondary electron emission in graphene under energetic ion irradiation, Phys. Rev. B 103 (2021) 224306. https://doi.org/10.1103/PhysRevB.103.224306.

[43] J. Cazaux, Calculated dependence of few-layer graphene on secondary electron emissions from SiC, Appl. Phys. Lett. 98 (2011). https://doi.org/10.1063/1.3534805.

[44] V.S. Fomenko, Handbook of Emission Properties of Materials, Kiev Nauk. Dumka (1981) 6.

[45] J. Pelletier, D. Gervais, C. Pomot, J. Pelletier, D. Gervais, C. Pomot, Application of widegap semiconductors to surface ionization : Work functions of AlN and SiC single crystals Application of wide-gap semiconductors to surface ionization : Work functions of AIN and SiC single crystals, 994 (2012). https://doi.org/10.1063/1.333156.

[46] M. Sabisch, P. Krüger, J. Pollmann, Ab initio calculations of structural and electronic properties of 6H-SiC(0001) surfaces, Phys. Rev. B 55 (1997) 10561–10570.

[47] P. Vashishta, R.K. Kalia, A. Nakano, J.P. Rino, Interaction potential for silicon carbide: A molecular dynamics study of elastic constants and vibrational density of states for crystalline and amorphous silicon carbide, J. Appl. Phys. 101 (2007). https://doi.org/10.1063/1.2724570.

[48] S.J. Zinkle, J.W. Jones, V.A. Skuratov, Microstructure of swift heavy ion irradiated SiC, Si3N4 and AlN, Mater. Res. Soc. Symp. - Proc. 650 (2001) 1–6. https://doi.org/10.1557/proc-650-r3.19.

[49] J.S. Alexander, C. Maxwell, J. Pencer, M. Saoudi, Equilibrium Molecular Dynamics Calculations of Thermal Conductivity: a “How-To” for the Beginners, CNL Nucl. Rev. 9 (2020) 11–25. https://doi.org/10.12943/cnr.2018.00009.

[50] R. Rymzhanov, N.A. Medvedev, A.E. Volkov, Damage threshold and structure of swift heavy ion tracks in Al2O3, J. Phys. D. Appl. Phys. 50 (2017) 475301. https://doi.org/10.1088/1361-6463/aa8ff5.

[51] M. Backman, M. Toulemonde, O.H. Pakarinen, N. Juslin, F. Djurabekova, K. Nordlund, A. Debelle, W.J. Weber, Molecular dynamics simulations of swift heavy ion induced defect recovery in SiC, Comput. Mater. Sci. 67 (2013) 261–265. https://doi.org/https://doi.org/10.1016/j.commatsci.2012.09.010.

[52] H.J.C. Berendsen, J.P.M. Postma, W.F. van Gunsteren, A. DiNola, J.R. Haak, Molecular dynamics with coupling to an external bath, J. Chem. Phys. 81 (1984) 3684–3690. https://doi.org/10.1063/1.448118.

[53] L. Zeng, X. Zhou, R. Cheng, X. Wang, J. Ren, Y. Lei, L. Ma, Y. Zhao, X. Zhang, Z. Xu, Temperature and energy effects on secondary electron emission from SiC ceramics induced by Xe17+ ions, Sci. Rep. 7 (2017) 1–6. https://doi.org/10.1038/s41598-017-06891-9.

[54] E.J. Sternglass, Theory of secondary electron emission by high-speed ions, Phys. Rev. 108 (1957) 1–12. https://doi.org/10.1103/PhysRev.108.1.

[55] Y. Zhang, R. Sachan, O.H. Pakarinen, M.F. Chisholm, P. Liu, H. Xue, W.J. Weber, Ionization-induced annealing of pre-existing defects in silicon carbide, Nat. Commun. 6 (2015) 1–7. https://doi.org/10.1038/ncomms9049.

[56] A. Debelle, M. Backman, L. Thomé, W.J. Weber, M. Toulemonde, S. Mylonas, A. Boulle, O.H. Pakarinen, N. Juslin, F. Djurabekova, K. Nordlund, F. Garrido, D. Chaussende, Combined experimental and computational study of the recrystallization process induced by electronic interactions of swift heavy ions with silicon carbide crystals, Phys. Rev. B 86 (2012) 100102. https://doi.org/10.1103/PhysRevB.86.100102.

[57] Z. Zhang, S. Jiang, X. Hu, J. Zhang, Study of the recrystallization behaviors induced by annealing and irradiation on amorphous SiC, J. Appl. Phys. 137 (2025) 015901. https://doi.org/10.1063/5.0232414.

[58] K.M. Tynyshbayeva, A.L. Kozlovskiy, R. V Rakhimov, V. V Uglov, M. V Zdorovets, Study of The Gas-Swelling Mechanisms in Silicon Carbide Ceramics under High-Temperature Irradiation with Helium Ions, Materials (Basel). 16 (2023) 5750. https://doi.org/10.3390/ma16175750.

[59] Z. Wang, L. Zhang, A. Tamer, B. Li, T. Polcar, N. Daghbouj, Exploring defect behavior in helium-irradiated single-crystal and nanocrystalline 3C-SiC at 800 ◦ C : A synergy of experimental and simulation techniques, Acta Mater. 279 (2024) 120281. https://doi.org/10.1016/j.actamat.2024.120281.

[60] A.S. El-Said, R.A. Wilhelm, R. Heller, M. Sorokin, S. Facsko, F. Aumayr, Tuning the Fabrication of Nanostructures by Low-Energy Highly Charged Ions, Phys. Rev. Lett. 117 (2016) 1–5. https://doi.org/10.1103/PhysRevLett.117.126101.

[61] O. Ochedowski, O. Osmani, M. Schade, B.K. Bussmann, B. Ban-d'Etat, H. Lebius, M. Schleberger, B. Ban-Detat, H. Lebius, M. Schleberger, Graphitic nanostripes in silicon carbide surfaces created by swift heavy ion irradiation, Nat. Commun. 5 (2014) 3913. https://doi.org/10.1038/ncomms4913.

[62] L.Q. Zhang, C.H. Zhang, C.C. Hao, G.D. Hu, Z.D. Feng, Y.J. Tang, W.K. Zhu, J.Y. Li, Q. Huang, Mechanical properties and surface characteristics of SiC fibers irradiated by swift heavy ions, Nucl. Sci. Tech. 33 (2022) 1–11. https://doi.org/10.1007/s41365-022-01095-x.

[63] N. Ishikawa, T. Taguchi, H. Ogawa, Comprehensive Understanding of Hillocks and Ion Tracks in Ceramics Irradiated with Swift Heavy Ions, Quantum Beam Sci. 2020, Vol. 4,

Page 43 4 (2020) 43. https://doi.org/10.3390/QUBS4040043.

[64] S. Nan, M. Xiao, Z. Guan, C. Feng, C. Huo, G. Li, P. Zhai, F. Zhang, Atomic-scale revealing defects in ion irradiated 4H-SiC, Mater. Charact. 203 (2023) 113125. https://doi.org/10.1016/j.matchar.2023.113125.

[65] A. Medvid, B. Berzina, L. Trinkler, L. Fedorenko, P. Lytvyn, N. Yusupov, T. Yamaguchi, L. Sirghi, M. Aoyama, Formation of nanostructure on surface of SiC by laser radiation, Phys. Status Solidi Appl. Res. 195 (2003) 199–203. https://doi.org/10.1002/pssa.200306281.

[66] X. Yan, P. Zhai, C. Yang, S. Zhao, S. Nan, P. Hu, T. Zhang, Q. Chen, L. Xu, Z. Li, J. Liu, Unveiling microstructural damage for leakage current degradation in SiC Schottky diode after heavy ions irradiation under 200 V, Appl. Phys. Lett. 125 (2024) 042103. https://doi.org/10.1063/5.0216883.

[67] N. Daghbouj, A.T. AlMotasem, J. Vesely, B.S. Li, H.S. Sen, M. Karlik, J. Lorinčík, F.F. Ge, L. Zhang, V. Krsjak, O. Laguta, M. Callisti, T. Polcar, Microstructure evolution of iron precipitates in (Fe, He)-irradiated 6H-SiC: A combined TEM and multiscale modeling, J. Nucl. Mater. 584 (2023) 154543. https://doi.org/10.1016/j.jnucmat.2023.154543.

[68] E. Matthias, M. Reichling, J. Siegel, O.W. Käding, S. Petzoldt, H. Skurk, P. Bizenberger, E. Neske, The Influence of Thermal Diffusion on Laser Ablation of Metal Films, Appl. Phys. A 58 (1994) 129–136. https://doi.org/10.1007/BF00332169.

[69] A.I. Karasevskii, V. V. Lubashenko, Size effect on thermodynamic properties of free nanocrystals, Eur. Phys. J. B (2008) 375–383. https://doi.org/10.1140/epjb/e2008-00429-9.

[70] Q. Jiang, Z. Wen, Thermodynamics of Materials, Springer, 2011.

[71] M. Takagi, Electron-Diffraction Study of Liquid-Solid Transition of Thin Metal Films, J. Phys. Soc. Japan 9 (1954) 359–363.

[72] A.P. Chernyshev, Temperature and Enthalpy of Fusion of Nanooxides UO2 and ThO2 of Different Morphologies, Russ. J. Phys. Chem. A 98 (2024) 2584–2591.

[73] X. Zhang, E. Kioupakis, Phonon-assisted optical absorption of SiC polytypes from first principles, Phys. Rev. B 107 (2023) 115207.

[74] S. Zhong, B. Liu, Y. Xia, J. Liu, J. Liu, Z. Shen, Z. Xu, C. Li, Solar Energy Materials & Solar Cells Influence of the texturing structure on the properties of black silicon solar cell, Sol. Energy Mater. Sol. Cells 108 (2013) 200–204. https://doi.org/10.1016/j.solmat.2012.10.001.

[75] Z. Huaiwen, Z. Yiyun, Y. Hua, X. Bin, W. Kui, L. Jing, W. Guohong, Light extraction efficiency enhancement in light- emitting diodes with indium tin oxide nano-craters, J. Semicond. 33 (2012) 054009. https://doi.org/10.1088/1674-4926/33/5/054009.

[76] H. Ou, Y. Ou, A. Argyraki, S. Kamiyama, V. Grivickas, S. Schimmel, M. Kaiser, P. Wellmann, M.K. Linnarsson, V. Jokubavicius, Light extraction efficiency enhancement for fluorescent SiC based white light-emitting diodes, in: 2nd Asian Symp. Electromagn. Photonics Eng. (ASEPE 2013), 2013.

[77] N. Medvedev, R. Rymzhanov, A. Volkov, TREKIS-3 [Computer Software], (2023) https://doi.org/10.5281/zenodo.8394462.

[78] A. Stukowski, Visualization and analysis of atomistic simulation data with OVITO–the Open Visualization Tool, Model. Simul. Mater. Sci. Eng. 18 (2010) 15012. https://doi.org/10.1088/0965-0393/18/1/015012.

[79] C.L. Kelchner, S.J. Plimpton, J.C. Hamilton, Dislocation nucleation and defect structure

during surface indentation, Phys. Rev. B 58 (1998) 85–88. https://doi.org/10.1103/PhysRevB.58.11085.

[80] J. Wang, L. Zhang, Q. Zeng, Surface relaxation and oxygen adsorption behavior of different SiC polytypes : a first-principles study, J. Phys. Condens. Matter 265003 (n.d.). https://doi.org/10.1088/0953-8984/22/26/265003.

[81] K. Daviau, K.K.M. Lee, High-pressure, high-temperature behavior of silicon carbide: A review, Crystals 8 (2018) 1–18. https://doi.org/10.3390/cryst8050217.

[82] I. Forbeaux, J. Themlin, J. Debever, High-temperature graphitization of the 6H-SiC ( 0001 ) face, Surf. Sci. 442 (1999) 9–18. https://doi.org/10.1016/s0039-6028(99)00891-2.

[83] E.J. Munoz, V. Attari, M. Martinez, M.B. Dickerson, M. Radovic, R. Arroyave, Phase-field model of silicon carbide growth during isothermal condition, Comput. Mater. Sci. 242 (2024) 113058. https://doi.org/10.1016/j.commatsci.2024.113058.

[84] F. Gao, W.J. Weber, Recovery of close Frenkel pairs produced by low energy recoils in SiC, 4348 (2014). https://doi.org/10.1063/1.1605254.